\documentclass[prx,superscriptaddress,aps,amsmath,amssymb,amsfonts,floatfix,reprint,raggedbottom,longbibliography]{revtex4-2}

\usepackage{graphicx}
\graphicspath{{Figures/}}
\usepackage{xr-hyper}
\usepackage[colorlinks=true,citecolor=blue,linkcolor=blue,urlcolor=blue]{hyperref}
\usepackage{balance,wrapfig}
\usepackage{dcolumn}
\usepackage{xcolor}
\usepackage{lipsum}
\usepackage{soul}
\usepackage{braket}
\usepackage{multirow}
\usepackage{array}
\usepackage{booktabs}
\usepackage[ruled,vlined,linesnumbered]{algorithm2e}
\usepackage{footnote}
\usepackage{titlesec}
\usepackage{lmodern}
\usepackage{times}
\usepackage[utf8]{inputenc}
\usepackage[subpreambles=true]{standalone}
\usepackage[para]{footmisc}
\usepackage[misc]{ifsym}

\DeclareUnicodeCharacter{2009}{\,}

\makeatletter
\newcommand*{\addFileDependency}[1]{
  \typeout{(#1)}
  \@addtofilelist{#1}
  \IfFileExists{#1}{}{\typeout{No file #1.}}
}
\makeatother

\makeatletter 
\renewcommand{\fnum@figure}{\textbf{Fig.~\thefigure}}
\makeatother

\makeatletter
\def\bbordermatrix#1{\begingroup \m@th
  \@tempdima 4.75\p@
  \setbox\z@\vbox{%
    \def\cr{\crcr\noalign{\kern2\p@\global\let\cr\endline}}%
    \ialign{$##$\hfil\kern2\p@\kern\@tempdima&\thinspace\hfil$##$\hfil
      &&\quad\hfil$##$\hfil\crcr
      \omit\strut\hfil\crcr\noalign{\kern-\baselineskip}%
      #1\crcr\omit\strut\cr}}%
  \setbox\tw@\vbox{\unvcopy\z@\global\setbox\@ne\lastbox}%
  \setbox\tw@\hbox{\unhbox\@ne\unskip\global\setbox\@ne\lastbox}%
  \setbox\tw@\hbox{$\kern\wd\@ne\kern-\@tempdima\left[\kern-\wd\@ne
    \global\setbox\@ne\vbox{\box\@ne\kern2\p@}%
    \vcenter{\kern-\ht\@ne\unvbox\z@\kern-\baselineskip}\,\right]$}%
  \null\;\vbox{\kern\ht\@ne\box\tw@}\endgroup}
\makeatother

\newcolumntype{L}{>{\arraybackslash}m{3.9 cm}}
\newcolumntype{C}{>{\centering\arraybackslash}m{3.2 cm}}
\newcolumntype{G}{>{\centering\arraybackslash}m{1.41 cm}}

\emergencystretch=\maxdimen
\begin{document}

\title{Pushing the Boundary of Quantum Advantage in Hard Combinatorial Optimization with Probabilistic Computers}
\author{Shuvro Chowdhury}
\affiliation{Department of Electrical and Computer Engineering, University of California, Santa Barbara, Santa Barbara, CA 93106, USA}
\author{Navid Anjum Aadit}
\affiliation{Department of Electrical and Computer Engineering, University of California, Santa Barbara, Santa Barbara, CA 93106, USA}
\author{Andrea Grimaldi}
\affiliation{Department of Mathematical and Computer Sciences, Physical Sciences and Earth Sciences, University of Messina, 98166, Messina, Italy}
\affiliation{Department of Electrical and Information Engineering, Politecnico di Bari, 70126 Bari, Italy}
\author{Eleonora Raimondo}
\affiliation{Department of Mathematical and Computer Sciences, Physical Sciences and Earth Sciences, University of Messina, 98166, Messina, Italy}
\affiliation{Istituto Nazionale di Geofisica e Vulcanologia, Via di Vigna Murata 605, 00143 Roma, Italy}  
\author{Atharva Raut}
\affiliation{Department of Electrical and Computer Engineering, Carnegie Mellon University}
\author{P. Aaron Lott}
\affiliation{USRA Research Institute for Advanced Computer Science (RIACS)}
\affiliation{Quantum Artificial Intelligence Laboratory (QuAIL), NASA Ames Research Center}
\author{Johan H. Mentink}
\affiliation{Institute for Molecules and Materials, Radboud University,  Heyendaalseweg 135, Nijmegen, The Netherlands}  
\author{Marek M. Rams}
\affiliation{Institute of Theoretical Physics, Jagiellonian University,  Lojasiewicza 11, PL-30348 Krak\'{o}w, Poland}
\author{Federico Ricci-Tersenghi}
\affiliation{Dipartimento di Fisica, Sapienza Universit\`a di Roma, and CNR-Nanotec,
Rome unit and INFN, Sezione di Roma 1, 00185 Rome, Italy}
\author{Massimo Chiappini} 
\affiliation{Istituto Nazionale di Geofisica e Vulcanologia, Via di Vigna Murata 605, 00143 Roma, Italy}  
\author{Luke S. Theogarajan}
\affiliation{Department of Electrical and Computer Engineering, University of California, Santa Barbara, Santa Barbara, CA 93106, USA}
\author{Tathagata Srimani}
\affiliation{Department of Electrical and Computer Engineering, Carnegie Mellon University}
\author{Giovanni Finocchio}
\affiliation{Department of Mathematical and Computer Sciences, Physical Sciences and Earth Sciences, University of Messina, 98166, Messina, Italy}
\author{Masoud Mohseni}
\affiliation{Emergent Machine Intelligence, Hewlett Packard Labs, CA, USA}
\author{Kerem Y. Camsari}
\affiliation{Department of Electrical and Computer Engineering, University of California, Santa Barbara, Santa Barbara, CA 93106, USA}
\date{\today}

\begin{abstract}
Recent demonstrations on specialized benchmarks have reignited excitement for quantum computers, yet whether they can deliver an advantage for practical real-world problems remains an open question. Here, we show that probabilistic computers (p-computers), when co-designed with hardware to implement powerful Monte Carlo algorithms, provide a compelling and scalable classical pathway for solving hard optimization problems. We focus on two key algorithms applied to 3D spin glasses: discrete-time simulated quantum annealing (DT-SQA) and adaptive parallel tempering (APT). We benchmark these methods against the performance of a leading quantum annealer on the same problem instances. For DT-SQA, we find that increasing the number of replicas improves residual energy scaling, in line with expectations from extreme value theory. We then show that APT, when supported by non-local isoenergetic cluster moves, exhibits a more favorable scaling and ultimately outperforms DT-SQA. We demonstrate these algorithms are readily implementable in modern hardware, projecting that custom Field Programmable Gate Arrays (FPGA) or specialized chips can leverage massive parallelism to accelerate these algorithms by orders of magnitude while drastically improving energy efficiency. Our results establish a new,  rigorous classical baseline, clarifying the landscape for assessing a practical quantum advantage and presenting p-computers as a scalable platform for real-world optimization challenges.
\end{abstract}

\maketitle

\twocolumngrid

\section{Introduction}
\label{sec:Intro}
Richard Feynman is widely credited with starting the field of quantum computing in   a 1982 lecture \cite{feynman1982simulating}. In the same  lecture, Feynman also introduced the notion of a probabilistic computer, one that naturally simulates probabilistic processes. Feynman's broader vision of \textit{physical computers}, or programmable physical devices that solve a problem of interest through their natural evolution, has recently inspired a growing array of physical and physics-inspired classical computing paradigms, including systems to train deep neural networks \cite{wright2022deep}, solve linear algebra problems \cite{aifer2024thermodynamic}, and tackle combinatorial optimization problems \cite{aadit2022massively}.

Building on this vision, a key challenge is identifying scenarios where scalable and error-corrected quantum computers  \cite{Mohseni2024} could outperform probabilistic or classical approaches, particularly in optimization and sampling tasks. Prominent examples include Shor's algorithm for factoring large integers \cite{shor1994algorithms}, sampling random quantum circuits \cite{arute2019quantum}, and learning quantum data on quantum processors \cite{Lloyd_2014,Huang_learningexperiments_2022}, each offering potential exponential speedups over all known probabilistic alternatives, typically due to the interference of probability amplitudes in a high-dimensional Hilbert space. However, the scaling challenges and quantum error correction overhead might diminish or eliminate such quantum advantages \cite{Mohseni2024}. Notably, while probabilistic computers can emulate quantum interference with polynomial resources, their convergence is in general believed to require exponential time \cite{chowdhury2023emulating}. This challenge is known as the sign-problem in Monte Carlo algorithms \cite{troyer2005computational}.

\begin{figure*}[!ht]
    \centering
    \vspace{0pt}
    \includegraphics[width=0.95\textwidth,keepaspectratio]{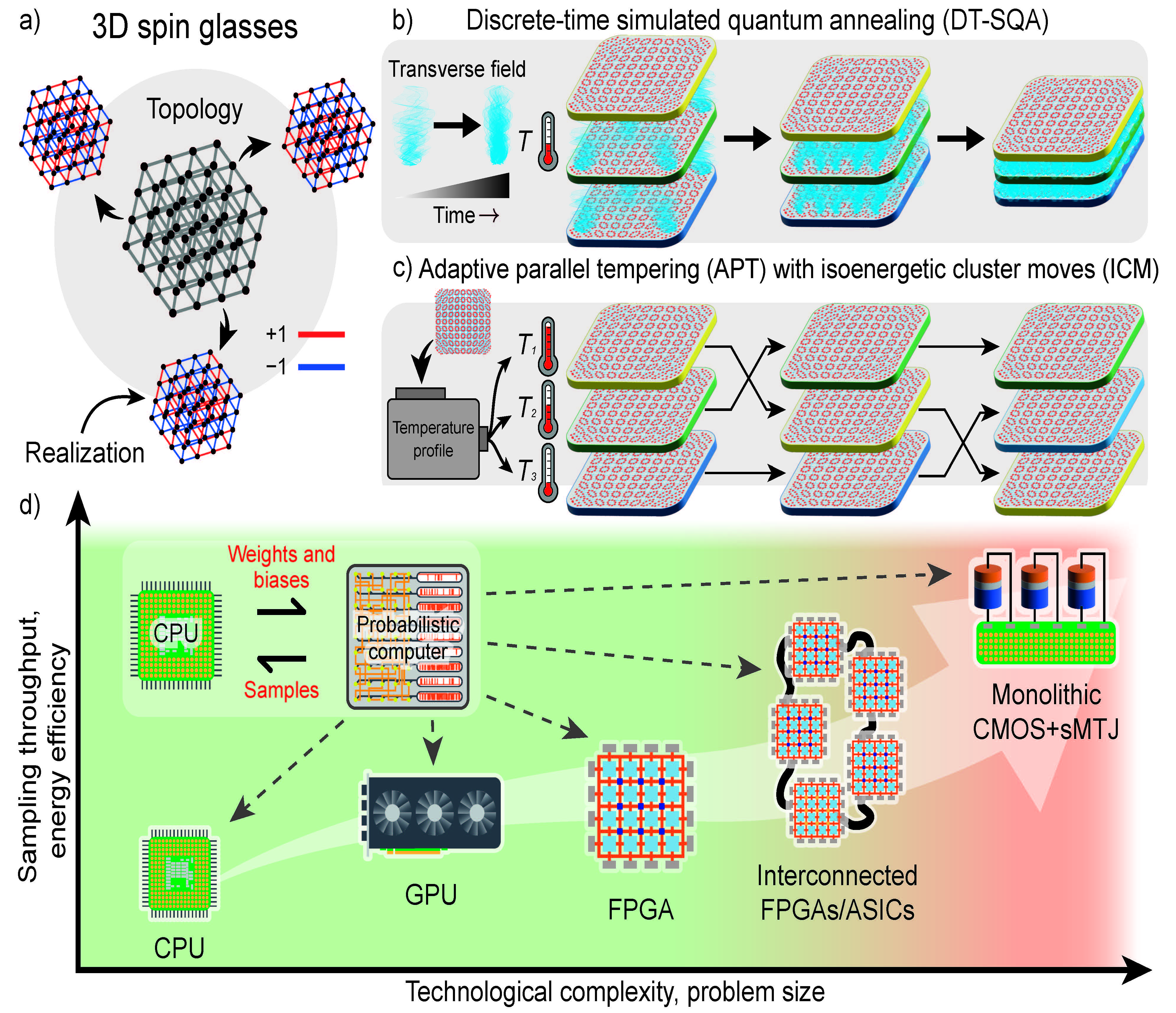}
    \caption{\footnotesize\textbf{Schematic overview of the probabilistic algorithms and technological platforms:} (a) Representation of a 3D Ising spin glass, where the weights connecting spins can be $+1$ (ferromagnetic, red) or $-1$ (antiferromagnetic, blue). Many real-world hard optimization problems can be mapped onto such spin-glass systems. (b, c) Two replica-based algorithms investigated in this work: discrete-time simulated quantum annealing (DT-SQA), in (b), and adaptive parallel tempering (APT) with isoenergetic cluster moves (ICM), in (c). In SQA, replicas are interconnected, and the strength of these interconnections is annealed according to a schedule. APT uses multiple replicas of the problem, each running at different temperatures, and periodically exchanges states between replicas to escape local minima. (d) The p-computing scheme, where a general-purpose CPU interfaces with a specialized probabilistic computer to efficiently implement Monte Carlo algorithms. Various implementations of probabilistic computers are shown, including CPU \cite{chowdhury2023accelerated, aadit2022massively}, GPU \cite{Onizawa2025,block2010multi, preis2009gpu, yang2019high, fang2014parallel}, Field Programmable Gate Arrays (FPGA) \cite{aadit2022massively, niazi2024training, nikhar2024all, aadit2023accelerating, aadit2021computing, aadit2022physics}, interconnected FPGAs, and monolithic CMOS + sMTJ (stochastic magnetic tunnel junction) chips \cite{singh2024cmos, grimaldi2022experimental, singh2023hardware, borders2019integer}. Each platform offers trade-offs in sampling throughput, energy efficiency, problem size, and technological complexity.}
    \label{fig:hardware_overview}
    \vspace{-5pt}
\end{figure*}

On the other hand, establishing a quantum advantage becomes much harder in cases where quantum fluctuations or quantum interference may be present, but not known to play a significant role. For example, even though quantum annealers by D-Wave operate on the transverse field Ising Hamiltonian, which does not suffer from the sign problem, empirical performance advantages have been sought to be demonstrated over the years \cite{king2021scaling, albash2018demonstration,denchev2016computational,hen2015probing}. In a similar attempt, Bernaschi et al. \cite{Bernaschi2024} clarified that for a 2D quantum spin glass, quantum annealing could still provide a speedup in entering the spin-glass phase under certain conditions. However, it is unclear whether these advantages extend to solving optimization problems and represent a fundamental improvement over classical algorithms, such as simulated quantum annealing (SQA) and adaptive parallel tempering (APT), or if they are primarily due to hardware acceleration. Speedups of this second type are also a feature of dedicated probabilistic computers when the hardware architecture is tailored for probabilistic algorithms \cite{chowdhury2023accelerated}.

Recently, King et al. \cite{King2023quantum} demonstrated another empirical scaling advantage over continuous-time simulated quantum annealing (CT-SQA) and simulated annealing (SA) in solving classical 3D cubic Ising spin glass problems (Fig.~\ref{fig:hardware_overview}(a)). Due to their hardness, 3D spin glasses have long served as canonical benchmarks for evaluating scaling behavior of various algorithms \cite{zhu2015efficient,fan2023searching,Lee2025}. 

The performance reported in Ref.~\cite{King2023quantum} provides a timely and valuable benchmark for the field. In this work, we use this benchmark to evaluate a powerful classical alternative: probabilistic computers co-designed with domain-specific hardware. While the quantum critical dynamics observed in that study are a significant physical finding, it is crucial to assess whether the resulting optimization performance is competitive with the most advanced classical techniques. Our study addresses this by demonstrating that p-computers, implementing state-of-the-art replica-based Monte Carlo algorithms (Fig.~\ref{fig:hardware_overview}(b, c)), can achieve a comparable, and in some cases more favorable, performance scaling on the same 3D spin glass problems. Specifically, we investigate discrete-time simulated quantum annealing (DT-SQA) and adaptive parallel tempering (APT) on various p-computer realizations (Fig.~\ref{fig:hardware_overview}(d)).

Our hybrid computing platform combines a general-purpose computer with a p-computer specializing in fast Monte Carlo sampling (Fig.~\ref{fig:hardware_overview}(d)). p-computers have been implemented in CPUs \cite{chowdhury2023accelerated, aadit2022massively}, GPUs \cite{Onizawa2025,block2010multi, preis2009gpu, yang2019high, fang2014parallel}, Field Programmable Gate Arrays (FPGAs) \cite{aadit2022massively, niazi2024training, nikhar2024all, aadit2023accelerating, aadit2021computing, aadit2022physics}, and interconnected FPGAs. Specialized accelerators using single and distributed FPGAs already provide orders of magnitude performance improvements over CPUs \cite{aadit2022massively}. Although small-scale p-computers using CMOS + stochastic magnetic tunnel junction technology (sMTJ) have been developed \cite{singh2024cmos}, monolithically integrated CMOS + sMTJ chips hold the greatest promise in terms of energy efficiency and performance. However, the large-scale monolithic integration of CMOS + sMTJ remains to be seen.

For our experiments, we use CPU  and FPGA implementations of p-computers. For scaling studies, we use CPUs when prefactors of solution times are  not critical and FPGAs when they are a priority.  Specifically, we use DT-SQA with a large number of physical replicas and select the best replica at the end of the annealing. Using extreme value theory, we relate scaling exponents to the number of replicas, achieving good agreement with our experiments. In addition, a powerful variant of PT, equipped with isoenergetic cluster moves (ICM)  \cite{Houdayer2001,Zhu2015, king2019quantum}, exhibits a transition from an initial gentler slope to a steeper one due to the non-local moves, providing superior scaling to DT-SQA. Finite-size scaling analysis reveals a collapse of residual energy curves, our primary metric of solution quality, for APT with the steeper slope emerging as a universal feature that delivers superior performance in large-scale optimization problems. This indicates superior performance in large-scale optimization problems, where minimizing the time-to-solution for a target residual energy is the key objective.  Projections based on open-source process design kits show that modern digital chip technology can accommodate a large number of on-chip replicas, making all of our algorithms readily manufacturable in single chips. We also analyze the prefactor and architectural improvements achievable through dedicated FPGA and ASIC implementations. The projections further extend to modern digital chips and CMOS + X-type architectures incorporating nanodevices.

\section{Residual Energy of 3D Spin Glasses}

The problem setting is the Edwards-Anderson spin glass on the 3D cubic lattice: 
\begin{align}
H = -\sum_{i<j}{J_{ij}\sigma_i\sigma_j}\;, 
\label{eq:EAmodel}
\end{align}
where $\sigma_i$ are Ising spins, $\sigma_i\in\{-1,+1\}$. The coupling weights $J_{ij}$ are non-zero exclusively for nearest‑neighbor pairs and, for those pairs, each $J_{ij}$ is randomly selected from $\{-1,+1\}$ with equal probability. One quantity of interest is the {residual energy} $\rho_\mathrm{E}^\mathrm{f}$ defined as a function of the annealing time $t_\mathrm{a}$: 
\begin{align}
\rho_\mathrm{E}^\mathrm{f}(t_\mathrm{a}) = \displaystyle{\frac{\langle E(t_\mathrm{a})-E_0\rangle}{n}}
\;,
\label{eq:rho_E_f}
\end{align}
where $E_0$ is the ground energy of the Hamiltonian $H$, $E(t_\mathrm{a})$ is the energy measured at the end of the annealing time $t_\mathrm{a}$ and $n$ is the number of spins in the system. The averaging is performed over different problem instances and multiple independent runs.

Experimental observations from probabilistic Monte Carlo algorithms and quantum annealers show that the residual energy scales as a power-law in $t_\mathrm{a}$: 
\begin{align}
\rho_\mathrm{E}^\mathrm{f} \propto t_\mathrm{a}^{-\kappa_\mathrm{f}} \;.
\label{eq:k_f}
\end{align}
where $\kappa_\mathrm{f}$ is the fitted scaling exponent describing the power-law decay of the residual energy.

While the performance scaling in Ref.~\cite{King2023quantum} is analyzed in the context of the Kibble-Zurek mechanism (KZM), it is noted there that the residual energy does not follow a simple prediction from critical dynamics alone, as also noted in Ref.~\cite{Bernaschi2024}. In our work, we are therefore primarily focused on the quality of solutions, using the residual energy scaling (Eq.~(\ref{eq:k_f})) simply as an empirical benchmark for different optimizers. We do not attempt to map our data onto specific KZM exponents, nor do we assume that near-critical power laws fully govern the eventual solution quality for these optimization problems.

\begin{figure*}[!ht]
    \centering
    \vspace{-10pt}
    \includegraphics[width=0.95\textwidth,keepaspectratio]{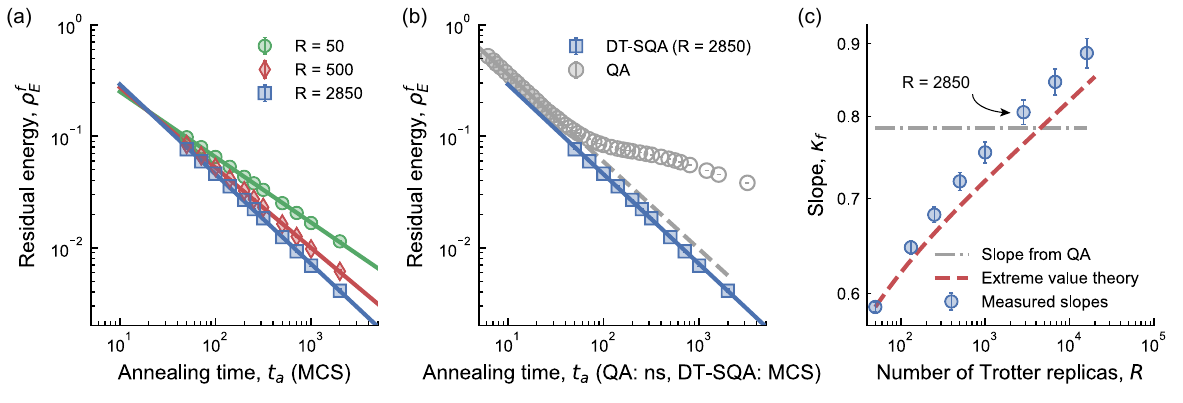}
    \caption{\footnotesize\textbf{Scaling improvement of the discrete-time simulated quantum annealing (DT-SQA) algorithm:} The final residual energy, $\rho_\mathrm{E}^\mathrm{f}$, as a function of annealing time, $t_\mathrm{a}$ (in Monte Carlo sweeps, (MCS)), for DT-SQA simulations with varying Trotter replicas, $R$, is shown for 3D Ising spin glass problems ($15 \times 15 \times 12$ with 2687 spins) using CPUs. The slope increases in absolute value with $R$. Each data point is averaged over 300 instances and 50 independent runs. The best energy among all Trotter replicas is selected at the end of $t_\mathrm{a}$. Error bars denote the 95\% bootstrap confidence interval of the mean across spin-glass instances. An MCS is an algorithmic unit representing one update attempt per spin; hardware differences alter the wall-clock time per sweep but not the scaling exponent $\kappa_{\mathrm{f}}$. The comparison with the quantum annealer's physical time in (b) is therefore a comparison of this dimensionless exponent, not a direct conversion of time units. A full analysis of wall-clock time, which incorporates these hardware-specific prefactors, is presented in Section~\ref{sec:arch_based} and Fig.~\ref{fig:prefactor}. (b) Comparison of the DT-SQA scaling exponent with that of the quantum annealer (QA) from Ref.~\cite{King2023quantum}. With a sufficient number of replicas ($R$\,$=$\,2850), DT-SQA achieves a more favorable scaling exponent. See Supplementary Fig.~\ref{fig:SQA_embedded2} for similar results on embedded instances. (c) The slope improvement is plotted against $R$, showing alignment with extreme value theory predictions (red dashed line; see Supplementary Section~\ref{sec:supp_evt}). The DT-SQA scaling exponent becomes comparable to the QA's at $R \approx 2800$ and exceeds it for larger values. Error bars denote 95\% confidence interval of fitting.}
    \label{fig:scaling_SQA}
    \vspace{0pt}
\end{figure*}

\section{Analysis of Residual Energy Scaling with DT-SQA}

The DT-SQA is an annealing-based algorithm inspired by the principle of adiabatically reducing the transverse field in a quantum system. Using the well-known Suzuki-Trotter transformation \cite{suzuki1976relationship}, a $d$-dimensional quantum system is mapped onto a ($d$$+$1)-dimensional classical system. The additional ``imaginary-time'' dimension is composed of $R$ interconnected Trotter replicas of the original quantum system, where qubits are replaced by Ising spins. As proposed in Ref.~\cite{camsari2019scalable}, our strategy is to implement the DT-SQA algorithm directly on probabilistic hardware, using distinct physical replicas. 

In Fig.~\ref{fig:scaling_SQA}(a), we evaluate the scaling performance of DT-SQA by plotting $\rho_\mathrm{E}^\mathrm{f}$ as a function of the annealing time, $t_\mathrm{a}$, with varying $R$. The inverse temperature is set to $\beta$~$=$~$0.5R$ and the simulations were performed on CPUs using logical problem instances defined on a 3D cubic lattice of Ising spins with dimensions $15 \times 15 \times 12 $, obtained directly from Ref.~\cite{King2023quantum}. The results show that the absolute value of the slope, $\kappa_\mathrm{f}$, increases with $R$ when the minimum energy among the $R$ replicas is selected. A comparison in Fig.~\ref{fig:scaling_SQA}(b) reveals that the scaling exponent of DT-SQA becomes comparable to that of the quantum annealer around $R$~$=$~$2850$ replicas (with $\kappa_\mathrm{f}$~$=$~$0.805$) and above. It is important to note that this is a comparison of the dimensionless scaling exponent, $\kappa_\mathrm{f}$, which is independent of the units on the time axis (MCS for p-computers, nanoseconds for the QA). The QA residual energy data is multiplied by a factor of 2 to align with logical instances, based on the observation that broken dimers (when physical spins representing the same logical spin do not agree after annealing) are rare \cite{King2023quantum}. In Ref.~\cite{King2023quantum},  $\kappa_\mathrm{f}$~$=$~$0.785$ and $\kappa_\mathrm{f}$~$=$~$0.51$ are quoted for the quantum annealer and CT-SQA algorithm, respectively. 

The slopes quoted above are based on embedded instances (logical problem instances mapped onto the quantum annealer's physical qubit connectivity graph). Quantum annealers (QAs) typically require complex embedding schemes for combinatorial optimization problems, in which a single logical spin is represented by multiple physical spins grouped into structures called dimers, due to their fixed hardware topology (such as the Chimera or Pegasus graphs), even for relatively sparse problems like 3D spin glasses. By contrast, probabilistic computers implemented on flexible classical hardware, such as FPGAs or ASICs, can directly represent and solve the logical problem graph without embedding overhead. Since our goal is to evaluate the intrinsic performance of algorithms solving practical combinatorial optimization problems, our main results do not include embedding overheads that are specific to current quantum annealer architectures.
Nevertheless, we essentially obtain similar results  also on embedded graphs (see Supplementary Fig.~\ref{fig:SQA_embedded2}) noting that the DT-SQA algorithm can match the scaling of quantum annealers in both cases.

Although both our work and Ref.~\cite{King2023quantum} employ SQA, our approach and goals are different. Ref.~\cite{King2023quantum} uses the continuous-time variant (CT-SQA) as a theoretical baseline, whereas we deliberately use the discrete-time SQA (DT-SQA) with $R$ explicit Trotter replicas. We chose DT-SQA as it maps naturally onto parallel hardware architectures, making it a more relevant algorithm for assessing the performance of physically realizable classical systems. Therefore, we benchmark our hardware-amenable algorithm directly against the quantum annealer's performance, rather than reproducing the CT-SQA baseline. In  Monte Carlo simulations intended to accurately emulate equilibrium quantum physics, selecting the best-performing replica among multiple Trotter replicas is usually avoided, as this could bias equilibrium observables~\cite{Heim2015quantum,Santoro2002}. However, such concerns are not relevant in our context, because our goal is not quantum emulation but rather practical combinatorial optimization. Indeed, since our replicas represent independent physical entities realized by separate physical spins in hardware, identifying and selecting the replica with the lowest residual energy is both natural and appropriate.

Next, we  show that the observed increase in slope $\kappa_\mathrm{f}$ with respect to $R$ can be explained using extreme value theory (EVT; see Supplementary Section~\ref{sec:supp_evt} for details) with modifications to account for correlations among replicas. In conventional EVT, the minimum energy is selected from $P$ independent runs of an algorithm, shifting the expected value of the minimum energy by $\mathcal{O}(\sqrt{\ln P})$ from the mean of the original distribution ($P$~=~1).  
In DT-SQA, the Trotter replicas are interconnected and  correlated, complicating a direct application of EVT. We observe however that the replica correlations decay over a distance, allowing us to extract effectively independent block sizes. Another complication is the dependence of the  correlations with $t_\mathrm{a}$, as the transverse coupling among the replicas ($J_\perp$) strengthens (see Supplementary Section~\ref{supp_sec:DTPIMC_algo} and Supplementary Fig.~\ref{fig:corr_study}). To apply an EVT theory despite these complications, we partition the $R$ Trotter replicas into $P$ effective blocks, where replicas within a block are correlated but largely uncorrelated with other blocks. We then treat these blocks as separate runs and observe that the predicted scaling behavior that aligns closely with the slopes observed in Fig.~\ref{fig:scaling_SQA}(c). The sizes of the  extracted effective blocks correspond well to the measured replica-to-replica correlations (Supplementary Fig.~\ref{fig:blkSize}(b)), providing further support for our modified EVT analysis. 

The modified EVT approach explains how increasing replicas within a single run improves the scaling. An alternative strategy, also based on EVT, is to leverage multiple independent runs: use a fixed number of interconnected Trotter replicas, run the algorithm $P$ times independently and then select the best energy from all runs. We find that by setting $R$~=~32 and running $P$~=~50 independent iterations (a total of 1600 replicas), followed by selecting the best solution, DT-SQA also achieves slopes comparable to those of the quantum annealer (see Supplementary Fig.~\ref{fig:approach2_results}). However, this approach remains valid over a shorter range of $t_\mathrm{a}$ before the power law breaks down and transitions to a flat plateau region (see Supplementary Fig.~\ref{fig:EVT_orig_comp}), showing that the two approaches are not equivalent. Nevertheless, both DT-SQA approaches--$R$~$=$~2850, $P$~$=$~1 (shown in Fig.~\ref{fig:scaling_SQA}) and $R$~$=$~32, $P$~$=$~50\,--are feasible for implementation on a single classical chip, as we discuss in Section~\ref{sec:chip_feasibility} where large groups of spins on the chip can be updated simultaneously with massive parallelism.

It is important to note that there is a fundamental difference between the two approaches we demonstrated. When $P$~$=$~$1$, all the DT-SQA replicas are connected and we perform a single run of the algorithm to match (and exceed as needed) QA slopes. The large number of replicas necessary to match the QA slopes is strong evidence of the efficiency of the quantum annealer, nonetheless, our results show that DT-SQA, when equipped with sufficient replicas, can achieve a comparable or more favorable scaling exponent. Naturally, the second approach where we pick the best results out of $P$~$>$~$1$ runs can also be applied to the quantum annealer to increase residual energy slopes, however, this does not change our main findings for $P$~$=$~$1$. 

Finally, we note that while $\kappa_\mathrm{f}$ is a useful metric, it may not fully capture the practical relevance of algorithms for large-scale optimization. As discussed in Section~\ref{sec:apt}, DT-SQA, despite exhibiting a steep power-law decay in residual energy at early times in instances defined on a fixed 3D spin glass lattice of size $15\times15\times12$, transitions into a plateau at longer annealing times, where the residual energy stagnates and shows little further improvement. It is ultimately outperformed by the APT algorithm, which is easier to implement and parallelize in hardware. APT achieves significantly lower residual energies with identical computational resources, showing that relying solely on $\kappa_\mathrm{f}$ as a performance metric can be misleading, as different algorithms may exhibit distinct scaling behaviors at different stages of optimization, which are crucial for real-world applications.

\begin{figure*}[!ht]
    \centering
    \vspace{0pt}
    \includegraphics[width=0.98\textwidth,keepaspectratio]{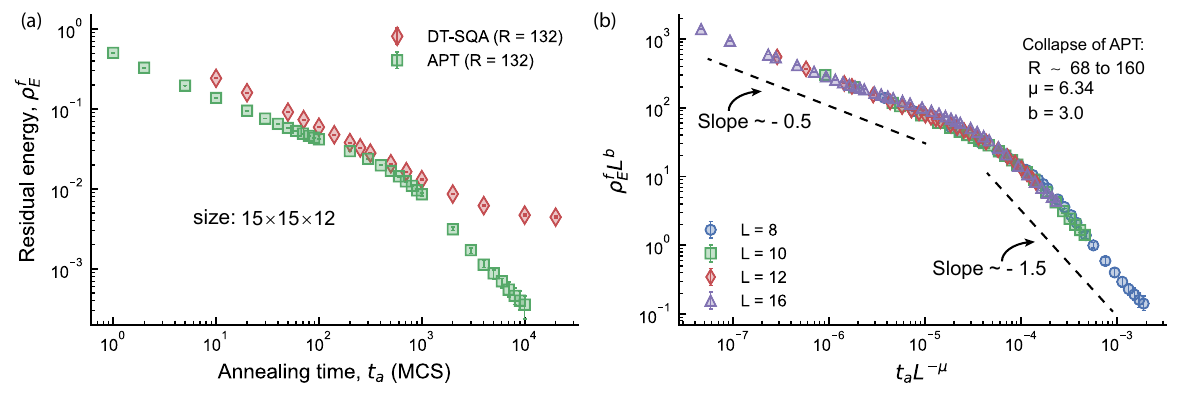}
    \caption{\footnotesize\textbf{Scaling improvement of the APT algorithm with ICM:}  (a) Iso-Trotter replica comparison of DT-SQA and adaptive parallel tempering (APT) with isoenergetic cluster moves (ICM) on CPUs for the same logical instances as in Fig.~\ref{fig:scaling_SQA}, using 50 independent runs per instance. A sweep-to-swap ratio of 1 minimizes residual energy for APT. While DT-SQA approaches a flat plateau, APT shows a sharp bending towards lower residual energy.
    (b) Finite-size scaling analysis for various cube sizes ($L$) reveals a universal collapse of the data. The later bending appears to be universal for APT, as shown by averaging over 200 independent runs per instance. For each size, 4 ICM replicas are used, with the total number of replicas varying due to the adaptive nature of the pre-processing algorithm. A similar collapse for APT without ICM is provided in the Supplementary Fig.~\ref{fig:rho_ta_L_woICM}. Error bars represent the 95\% bootstrap confidence interval of the mean across spin-glass instances.}
    \label{fig:prefactor_plot}
    \vspace{0pt}
\end{figure*}

\section{Comparison with Adaptive Parallel Tempering}
\label{sec:apt}
We now compare DT-SQA with the powerful replica-based adaptive parallel tempering (APT) algorithm, widely considered as the state-of-the-art for solving spin-glass problems \cite{Billoire_2018, Papakonstantinou_2014, earl2005parallel, swendsen1986replica, hukushima1996exchange, andrea2022spintronics, Isakov2015optimising}. APT also utilizes replicas of the problem graph, but these run in parallel at different temperatures, with adjacent replicas periodically swapped based on the Metropolis criterion:
\begin{align}
p_{\text{swap}} = \mathrm{min}\left(1, \exp{\left(\Delta\beta\Delta E\right)}\right), 
\end{align}
where \(\Delta E\)~$=$~\(E_{i+1} - E_{i}\) is the energy difference, and \(\Delta \beta = \beta_{i+1} - \beta_{i}\) is the difference in inverse temperatures between replicas \(i\) and \(i\)$+$\(1\) (with \(\beta_i\)~$<$~\(\beta_{i+1}\)). This mechanism enables high-temperature replicas to explore the energy landscape broadly while low-temperature replicas preserve optimal states. The adaptive variant further optimizes the algorithm by preprocessing the problem graph to equalize swap probabilities across replicas, avoiding bottlenecks \cite{Katzgraber_2006, Isakov2015optimising}.

Fig.~\ref{fig:prefactor_plot}(a) compares DT-SQA and APT for the same problem as in Fig.~\ref{fig:scaling_SQA}. Adaptive preprocessing produces approximately 33 temperature replicas per instance (see Section~\ref{sec:methods}). To further enhance APT, we incorporate isoenergetic cluster moves (ICM) \cite{Houdayer2001,Zhu2015, king2019quantum}, which, as we demonstrate later, play a crucial role. ICM are non-local Monte Carlo swaps added on top of the standard APT algorithm. A swap attempt follows each network sweep, maintaining a sweep-to-swap ratio of 1. Using 4 replicas per temperature for ICM, the APT algorithm used in this work operates with a total of 132 replicas. As shown in Supplementary Fig.~\ref{fig:ICM_sweep_swap}, we found that this sweep-to-swap ratio produces the smallest residual energy for a fixed MCS budget.

Optimization of APT parameters (detailed in Supplementary Section~\ref{supp_sec:APT_algo}) reveals that the initial slope of the optimized APT with ICM corresponds to $\kappa_\mathrm{f}$~$=$~$0.53$, slightly lower than DT-SQA with a similar number of replicas ($\kappa_\mathrm{f}$~$=$~$0.647$ at $R$~$=$~$132$). However, APT achieves lower residual energy for a given MCS budget. Although DT-SQA initially shows a better slope, it plateaus at higher MCS, as shown in Fig.~\ref{fig:prefactor_plot}(a). This trend is  observed across various cube sizes $L$ and Trotter replicas $R$ (see Supplementary Fig.~\ref{fig:DTSQA_L_R}) and  is consistent with previous findings \cite{Heim2015quantum}. In contrast, APT with ICM shows two distinct scaling regimes: an initial gentler slope followed by a steeper one (see Supplementary Figs.~\ref{fig:APTvsICM}(a) and \ref{fig:Houdayer_effect}). Notably, the APT algorithm without ICM does not exhibit this steeper bending, even with the same number of replicas (see Supplementary Fig.~\ref{fig:APTvsICM}(a) and Supplementary Fig.~\ref{fig:rho_ta_L_woICM}). The presence of this bending suggests the potential for algorithms that incorporate non-local and non-equilibrium moves \cite{Mohseni2021} to further enhance the performance of probabilistic approaches in solving hard optimization problems. As before, a similar performance characteristic is observed for embedded instances (see Supplementary Fig.~\ref{fig:APT_embedded}).

This steeper slope is also observed for other cube sizes $L$ (see Supplementary Fig.~\ref{fig:rho_ta_L}(a)) and appears to be a universal feature of APT supplemented by ICM. Finite‐size scaling analysis confirms that the residual‐energy curves for different sizes collapse onto a single universal curve (Fig.~\ref{fig:prefactor_plot}(b)). However, at very low residual energies near the ground state, we observe another transition to a gentler slope (not visible in Fig.~\ref{fig:prefactor_plot}(b)). This transition occurs at residual energies that are very close to the uncertainty limit of the ground energy estimations used in our analysis (see Supplementary Fig.~\ref{fig:rho_ta_L}(b)). As such, it is difficult to reliably confirm the existence of this feature. 
On the other hand, the robust universal collapse shown in Fig.~\ref{fig:prefactor_plot}(b) allows us to extrapolate the time required to reach a target residual energy for cubes of any size. 

\begin{figure*}[!ht]
    \centering
    \vspace{0pt}
    \includegraphics[width=0.98\textwidth,keepaspectratio]{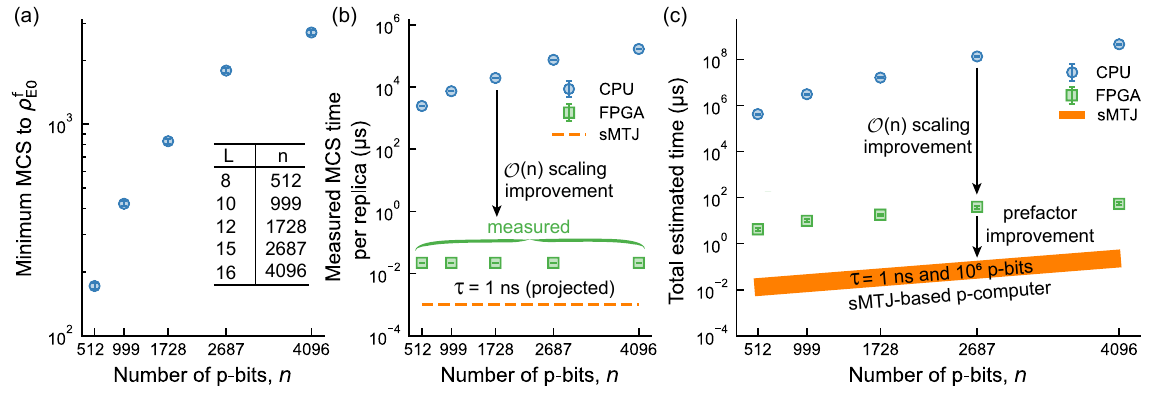}
    \caption{\footnotesize\textbf{Architectural improvement with probabilistic computers in hardware:} (a) Algorithmic complexity of the 3D spin glass problem as a function of the number of p-bits, showing the minimum number of Monte Carlo sweeps (MCS) required to reach a target threshold residual energy, $\rho_{\mathrm{E0}}^\mathrm{f} = 0.007$, using APT + ICM, independent of hardware platform. For each lattice size ($L$), 300 logical instances with 50 independent runs are reported. Swaps are probabilistically performed pairwise between adjacent $\beta$ replicas, alternating between even and odd pairs. Error bars (95\% confidence) are small and often invisible. (b) Measured average MCS time per replica for CPU and FPGA. CPU shows $\mathcal{O}(n)$ scaling, while FPGA achieves $\mathcal{O}(1)$ scaling due to the massively parallel p-computer architecture. sMTJ projections assume experimentally demonstrated nanosecond p-bits \cite{hayakawa2021nanosecond, soumah2024nanosecond}. (c) Total estimated time to reach $\rho_{\mathrm{E0}}^\mathrm{f} = 0.007$, combining (a) and (b). Time for ICM and swaps are negligible compared to sampling times (see Methods section) and are excluded from this estimation. Also, it is assumed that FPGA can run all the replicas in parallel (as long as they fit on a single chip \cite{nikhar2024all}; see Supplementary Section~\ref{sec:supp_chip}). We assume the same for CPU as well and do not include the replica factor when estimating these times. FPGA maintains $\mathcal{O}(n)$ improvement over CPU, while `prefactor improvement' for sMTJ-based p-computers refers to the additional constant speed-up expected when the same architecture is implemented with fast, on-chip sMTJ p-bits (approximately 1 ns intrinsic flip time), relative to the measured FPGA sweep time. sMTJ projections assume a single chip with 1 million sMTJs, achieving 1 million flips per ns (the thick orange line assumes 10 and 50 replicas for the upper and lower bounds, respectively). This also includes improvements stemming from additional parallelization, meaning that if the problem size is smaller than the chip's capacity, multiple independent runs or problem instances can be processed simultaneously on the same chip,  scaling as $10^6 / (\text{spins per replica} \times \text{APT replicas})$.}
    \vspace{0pt}
\label{fig:prefactor}
\vspace{0pt}
\end{figure*}

\section{Architecture-based scaling improvement and massive parallelism}
\label{sec:arch_based}

Beyond scaling improvements, a critical metric for optimization problems is achieving the lowest possible residual energy within a given amount of time. Here, we evaluate the relative performance of CPU, FPGA (see Supplementary Section~\ref{supp_sec:FPGA_verif} for the details of FPGA implementation), and sMTJ-based implementations of p-computers, highlighting architectural advantages. One key feature of the p-computer architecture we adopt here is the ability to probabilistically update all spins in the system in constant time. This differs from the sequential or partially parallel updates typically used in software implementations. In sparse problems, such as 3D spin-glasses, planted Ising benchmarks \cite{hamze2018near,hen2015probing} or circuit SAT problems with sparse connectivity \cite{aadit2022massively}, p-computer architectures  leverage physically parallel nodes to simultaneously update large independent sets \cite{aadit2022massively}. Fig.~\ref{fig:prefactor}(a) shows the number of Monte Carlo sweeps (MCS: one MCS involves one update attempt per spin for all spins in the network) needed to reach a target residual energy threshold of $\rho_\mathrm{E0}^\mathrm{f}$ = $0.007$, an arbitrarily chosen optimization goal given a computational budget. Using the APT + ICM algorithm, we show the required number of MCS as a function of the lattice size $L$. Fig.~\ref{fig:prefactor}(b) compares the time required for one sweep per replica across three architectures: CPU, FPGA-based p-computers, and sMTJ-based p-computers. While sweep times increase with problem size on CPUs, they remain constant, $\mathcal{O}(1)$, for FPGA- and sMTJ-based p-computers, exploiting massive parallelism until resource limits are reached. For the problem considered, $n/2$ or $n/4$ spins (depending on whether logical or embedded instances are used) can be updated simultaneously, resulting in an $\mathcal{O}(n)$ performance improvement over CPUs. The current FPGA implementation yields up to 185 flips/ns approaching performance of the state-of-the-art GPUs and FPGA-based simulators \cite{BAITYJESI2014550, Bernaschi2024}, with further possible improvements using specialized ASICs and nanodevices. Note however, that the p-computers we propose here can support arbitrary sparse graph topologies beyond the regular and more easily parallelizable topologies shown in \cite{BAITYJESI2014550,Bernaschi2024}. Fig.~\ref{fig:prefactor}(c) combines results from (a) and (b), showing total runtime as the product of sweep count and average sweep time. The analysis confirms an $\mathcal{O}(n)$ scaling advantage for p-computers over CPUs. Furthermore, sMTJ-based devices could achieve an additional 1 to 3 orders of magnitude improvement in prefactors, assuming nanosecond fluctuations in a 1-million p-bit MRAM chip, which are feasible.

The proposed architecture is also highly energy efficient, consuming 2 to 5 orders of magnitude less energy per flip compared to the state-of-the-art GPUs and TPUs used for probabilistic tasks. Our FPGA implementation consumes 9.168 W, which corresponds to  \(5 \times 10^{-2}\) nJ/flip. sMTJ-based devices, with 1 million p-bits integrated on a single MRAM chip, reduce this further to \(2 \times 10^{-5}\) nJ/flip assuming 20 W power consumption \cite{sutton2020autonomous}. In comparison, a NVIDIA Tesla V100 GPU consumes 21.99 nJ/flip, and a Google TPU v3 requires 7.77 nJ/flip to solve probabilistic problems on simpler graphs \cite{yang2019high}.

There is  a trade-off between reconfigurable and application-specific hardware. FPGAs offer full reconfigurability, ideal for algorithmic exploration across diverse problem structures, but at a significant performance and energy cost. For the 3D spin glasses studied here, the fixed nearest-neighbor topology is an excellent match for a custom ASIC. Connectivity can be hard-wired while programmability for different instances is retained by reloading different weights. For problems with arbitrary sparse topologies, however, achieving reconfigurability on static ASICs is an open problem and may require different approaches, such as higher-order problem formulations or master graph approaches ~\cite{iyer2025efficient,nikhar2024all}.

\section{Physical design feasibility of single p-computing chips}
\label{sec:chip_feasibility}

We now evaluate the feasibility of a custom Application-Specific Integrated Circuit (ASIC) designed for replica-based algorithms on sparse, structured problems. A monolithic chip that can house all replicas on-die would eliminate the off-chip communication overhead that constrains current FPGA implementations. To make realistic projections for a full-scale ASIC, our analysis is grounded in a rigorous, bottom-up physical design flow using a 7 nm process. Our findings on chip capacity for the DT-SQA algorithm apply equally to the better performing APT algorithm, which requires significantly fewer replicas, as shown in Section~\ref{sec:apt}.

The details of the p-computer architecture are discussed in Supplementary Section~\ref{sec:supp_chip}.  As shown in Supplementary Fig.~\ref{fig:chip_area} and Supplementary Table~\ref{tab:mflowgen_results}, a full place-and-route analysis was performed using the ASAP7 7 nm open-source process design kit (PDK) \cite{asap7} for up to 5 replicas of the $15\times15\times12$ logical instances. The analysis revealed an approximately linear growth (with a slope of 1.05) in chip area requirements.

Based on this scaling, we project that approximately 7.66 million p-bits -- corresponding to 2850 replicas can  fit on a single chip using 7 nm technology. This translates to a chip area of $28.61\times28.61$ mm$^2$, which is within the capabilities of current fabrication technology. Furthermore, multiple such chips can be interconnected to support even larger numbers of p-bits as needed. With advances in fabrication technology and the adoption of nanodevice-based p-bits, the number of p-bits per chip can be significantly increased, enabling even greater scalability.

\section{Outlook}

This paper demonstrates that probabilistic computing with p-bits provides a practical and scalable approach to solving 3D Ising spin glass problems. Using the DT-SQA algorithm, we showed how leveraging a large numbers of replicas greatly improves scaling exponents, well-explained by extreme value theory. We further explored  powerful algorithms like APT supported with non-local moves, significantly improving scaling and time-to-solution. Finite-size scaling analysis  revealed a universal collapse of residual energy curves for APT, emphasizing the generality of our results.  This makes APT particularly well-suited for large-scale optimization tasks when implemented on dedicated probabilistic hardware, as demonstrated by our FPGA-based implementation, achieving high performance through hardware acceleration.  

Advances in fabrication technology now allow large-scale replica systems, delivering orders-of-magnitude speedups compared to software methods. 
Projections for monolithic nanodevice-based p-computers highlight a path toward performance competitive with and potentially exceeding current solvers, all while operating at room temperature and without the specific hardware challenges of qubit decoherence or fixed-connectivity embedding.  Co-designed together, powerful algorithms, scalable architectures, and emerging hardware provide a clear pathway for solving hard optimization problems at unprecedented scales. Beyond combinatorial optimization, probabilistic computers hold promise for diverse applications including training and inference in energy-based models and Bayesian learning and in general for sampling over discrete spaces where the performance of traditional solvers have saturated. 

\section{Methods}
\label{sec:methods}

\subsection{p-computing overview}
\label{subsec:p-computing_overview}
p-computing relies on an interacting network of p-bits ${\sigma_i}$, which generate two-valued outputs ($\sigma_i \in \{-1, +1\}$) and are governed by two key equations \cite{camsari2017stochastic}:
\begin{align}
I_i = \sum_{j}J_{ij}\sigma_j+h_i \label{eq:synapse}\\
\sigma_i = \text{sgn}\left(\tanh(\beta I_i)-r_{[-1,1]}\right)
\label{eq:neuron}
\end{align}
Here, $J$, $h$, and $\beta$ represent the interconnection matrix, bias vector, and inverse temperature, respectively. $r_{[-1,1]}$ is a random number uniformly distributed in the range $[-1, 1]$. Equations (\ref{eq:synapse}) and (\ref{eq:neuron}) collectively approximate the Boltzmann distribution:
\begin{align}
p(\{\sigma_i\}) = \cfrac{1}{Z}\exp{\left(-\beta E(\{\sigma_i\})\right)}\label{eq:boltz_prob}\\
E(\{\sigma_i\}) =  -\sum_{i<j}{J_{ij}\sigma_i\sigma_j}-\sum_{i}h_i\sigma_i
\label{eq:ising_energy}
\end{align}
where \(p(\{\sigma_i\})\) represents the probability and \(E(\{\sigma_i\})\) represents the energy of the state \(\{\sigma_i\}\), with \(Z\) as the partition function.

\subsection{Instances}
For comparison, we use the instances from Ref.~\cite{King2023quantum} for $L$~$=$~$8,10$, and instances of size $15\times15\times12$. These instances have open boundaries along the $x$ and $y$ directions and periodic boundaries along the $z$ direction. For sizes greater than $L$~$=$~$9$, some spins are missing due to embedding constraints of the quantum annealer. Consequently, the total number of qubits for $L$~$=$~$10$, for example, is 
999, instead of the expected $L^3$~$=$~$1000$. We also use the putative ground energies reported in Ref.~\cite{King2023quantum} for these sizes. However, for a few instances, we found lower ground state energies than those reported and therefore used the improved ground energies in our analysis.

For $L$~$=$~$12$ and $16$, we generate instances using the codes provided in Ref.~\cite{King2023quantum}, ensuring that these instances do not suffer from missing spins. For these instances, the putative ground state energies are obtained by running APT with ICM algorithm up to $10^7$ sweeps (in each sweep all replicas are updated once), in a single run with a sweep-to-swap ratio of 10, choosing the minimum energy found along the whole simulation, and following a similar fitting and limiting procedure discussed in \cite{King2023quantum}. Our corresponding approximate error estimate per site for these sizes,  is $2.5\times10^{-4}$ (attributed to the increased problem size and the use of a single run) as indicated in Supplementary Fig.~\ref{fig:rho_ta_L}. This is comparable to the estimated error in the mean ground state energy per site, $4\times10^{-5}$, as reported in Ref.~\cite{King2023quantum} for lattice size $15\times15\times12$. The  residual energy ranges used in this work  to draw our conclusions are well above the range of these errors, or otherwise carefully discussed.

\subsection{Graph Coloring of 3D Cubic Spin Glass Instances}

If two p-bits in a network are not connected, they can be updated in parallel \cite{aadit2022massively}. Graph coloring assigns colors to the network such that any two connected p-bits are given different colors, while p-bits that are not connected can share the same color. This enables massive parallelism for sparse graphs even if they are irregular, allowing a network of p-bits to be updated in constant time, regardless of network size.

A perfect 3D lattice is bipartite and easily 2-colorable. However, the D-Wave instances have missing spins and complex embeddings (due to hardware constraints), which necessitates graph coloring. In this work, graph coloring is performed using DSATUR \cite{Brelaz1979}, a heuristic graph coloring algorithm with polynomial-time complexity. Since the underlying graph is identical for all problem instances of a given size, we perform graph coloring for one representative instance of each size as a preprocessing step. These problem instances typically require 2 to 4 colors, depending on their connectivity. In DT-SQA, replicas are connected and periodic boundary condition is applied. As a result, networks with odd numbers of replicas require an extra color.

\subsection{Annealing Schedule of DT-SQA}
Supplementary Section~\ref{supp_sec:DTPIMC_algo} details the description of the DT-SQA algorithm. Annealing is performed by gradually changing the transverse field ($\Gamma_x$) from a high value to 0. Change in $\Gamma_x$ is reflected in the coupling strength $J_{\perp}$ (see Supplementary Eq.~(\ref{supp_eq:SQA_ham})), which couples the spins of two neighboring replicas. 

In our implementation, we use a slightly modified form for $J_\perp$:
\begin{align}
J_\perp(t) = -\displaystyle{\frac{1}{\beta}\ln{\tanh{\left(\frac{\beta\,\Gamma'_x(t)}{R}\right)}}}
\end{align}
and anneal $\Gamma'_x(t)$ linearly, from 3.0 to 0. This modification represents a transformation between 
$\Gamma'_x$ and $\Gamma_x$ and does not alter the underlying physics. We also set $\beta/R = 0.5$ in all our simulations.

\subsection{APT details}
For the APT algorithm, we start with a preprocessing step to compute the inverse temperature ($\beta$) schedule and determine the required number of replicas. We perform the preprocessing individually for each of the 300 instances, even though schedules and number of replicas obtained are similar (see Supplementary Section~\ref{supp:beta_profiles}). Specific details of the preprocessing algorithm we adopted can be found in \cite{Mohseni2021, nikhar2024all} and Supplementary Algorithm \ref{algo:APT}. For our simulations, we set the initial inverse temperature to \(\beta_0 = 0.5\) and the temperature update factor to \(\alpha = 1.25\). We calculate the average energy variance across 100 parallel chains, where the variance for each chain is computed from the last 1000 sweeps of a 10000-sweep run before updating the temperature schedule. This process is repeated until the average energy variance drops below \(\text{min}(J_{ij})/2\). For the 300 instances with lattice size \(15\times15\times12\), the number of replicas ranged from 32 to 34. 

After determining the \(\beta\) schedule, each instance is simulated using the parallel tempering algorithm, both with and without the isoenergetic cluster moves (ICM) \cite{Houdayer2001,Zhu2015,king2019quantum}. We employ 4 ICM replicas per temperature. During simulation, each replica undergoes a fixed number of sweeps before a swap is attempted. A swap attempt involves performing an isoenergetic cluster move for each of the two randomly chosen ICM replica pairs at each temperature. This is followed by a swap attempt between neighboring replicas, determined as follows: for an odd-numbered swap attempt, pairs \((1,2), (3,4), \ldots\) are swapped; for an even-numbered swap attempt, pairs \((2,3), (4,5), \ldots\) are swapped. Within a given \(\beta\), ICM replicas are labeled \(a, b, c, d\), and swaps between neighboring \(\beta\) values occur only between replicas with the same label. The algorithm is detailed in Supplementary Algorithm \ref{algo:APT}.

\subsection{APT collapse}
The collapse of APT with ICM residual energies was obtained using the open-source library autoScale.py \cite{melchert2009}. The parameter $b$ fluctuates slightly around 3.0; we use $b = 3.0$, as it intuitively reflects the fact that the residual energy behaves extensively, scaling with the system size, $L^3$.

\subsection{CPU details}
All CPU-based simulations were run on a 10-core Intel Core i9-10900 processor (2.80 GHz) with 64 GB RAM, using MATLAB R2023b on a 64-bit Windows 10 machine. p-bits were updated sequentially using Gibbs sampling. All computations used MATLAB’s default double-precision arithmetic and Mersenne Twister pseudorandom number generator (PRNG).

\subsection{FPGA details}
We mapped the physics‑inspired, massively parallel \textit{p}-computer architecture of Ref.~\cite{aadit2022massively} onto a Xilinx Alveo U250 data‑center accelerator card using graph coloring to maximize parallelism on the sparse instances.     All arithmetic is fixed‑point: DT‑SQA uses s\(\{6\}\{3\}\) precision (1 sign, 6 integer, 3 fractional bits) while APT + ICM uses the higher s\(\{6\}\{6\}\) precision.  Custom RTLs were developed to implement the algorithm based on the p-computing architecture and synthesized, placed and routed with Xilinx Vivado/Vitis tool chain. Further details are provided in Supplementary Sections~III (DT‑SQA) and~V (APT). 

\subsection{FPGA implementation of APT with ICM algorithm}
At this time, we can accommodate only 1 replica in a single FPGA for large scales, such as instances of size $15\times15\times12$. To implement APT for this size, 32 to 34 replicas are required, while APT with ICM requires 128 to 136 replicas. We address this limitation by employing time-division multiplexing (TDM), allowing the same hardware to be reused for multiple replicas. At the start of each run, the weights of all replicas (scaled by $\beta$) are loaded into the BRAM. During each sweep, based on the replica index that will be sampled next, the weights corresponding to that replica are dynamically fetched from the BRAM. Then after the sampling is done, the state of p-bits of the current replica are also stored in the BRAM. This process is repeated until all replicas are sampled once. Then from MATLAB, we read all the p-bits states, (perform ICM whenever applicable) compute energies and perform the swaps. For the subsequent swap attempts, replicas are reinitialized either from the state saved in the previous swap or from the new biases (to restore certain p-bits to their original states before hardware was reused) generated by the APT swap/ICM. After initialization, the biases are released, and the p-bits resume their usual MCS at their respective $\beta$.

\subsection{Measurement of flips per second and time per MCS}
To measure the time per MCS in the FPGA accurately, we implemented precise counters within the FPGA to track the number of flip attempts made by each p-bit during a fixed time interval. This reference interval is determined by a predefined counter running at 125 MHz, which counts up to 50000. All counters are simultaneously enabled by a global signal from MATLAB and stop when the reference counter completes its count. This corresponds to an elapsed time of  $50000 / (125 \times 10^6)$  seconds (= 400 $\mathrm{\mu}$s). The total flip attempts during this period are summed across all p-bit counters to compute the total flips per second (fps).

Since one MCS involves updating all p-bits in a single replica, the time per MCS is calculated using the total measured flips and the elapsed time. For each network size  $L$, we performed 100 measurements and reported the average time per MCS. FPGA measurements are instance-independent. We reported MCS times for a single replica, given our detailed feasibility analysis that shows all relevant replicas sizes we considered can fit on a single chip. This full integration would eliminate the overhead for time-division multiplexing. It should be noted that, at present, the overhead from swapping and ICM moves dominates the MCS time because it is performed off-chip on the CPU. This is not a fundamental limitation: the overhead can be computed directly on the FPGA rather than off-chip on the CPU. Using standard hardware design flows, such as those enabled by high‐level synthesis (HLS), these computations can be seamlessly implemented on-chip, thus making their contribution negligible compared to the MCS time. Consequently, it was excluded from the FPGA measurements. We emphasize that none of the overheads that are omitted here affect our scaling exponent measurements where prefactors in time per MCS does not affect slopes in power laws.

For CPU measurements, MATLAB’s built-in \textit{tic} and \textit{toc} functions were used to measure the time taken to perform 100 MCS across 10 instances with 10 runs each. The average time per MCS is reported for a single replica of each network size $L$. Swap times were similarly excluded from CPU measurements to ensure a fair comparison with FPGA performance. For both FPGA and CPU measurements, error bars represent 95\% confidence intervals and were computed using bootstrapping with replacement.

\section*{DATA AVAILABILITY}
The data that support the findings of this study are available from the corresponding author upon reasonable request. 

\section*{CODE AVAILABILITY}

Simplified MATLAB implementations of the SQA and APT~+~ICM algorithms and the instances used in this study are openly available at
\href{https://github.com/OPUSLab/3DSpinGlassWithPbits.git}{https://github.com/OPUSLab/3DSpinGlassWithPbits.git}.
Pseudocode for both routines is provided in Supplementary Algorithms~\ref{algo:DTPIMC} and~\ref{algo:APT}.

\section*{Acknowledgments}
SC, KYC, and NAA acknowledge support from the Office of Naval Research (ONR) Young Investigator Program grant, the National Science Foundation (NSF) CAREER Award under grant number CCF 2106260, the Army Research Laboratory under grant number W911NF-24-1-0228, the Semiconductor Research Corporation (SRC) grant, and the ONR-MURI grant N000142312708. Use was made of computational facilities purchased with funds from the National Science Foundation (CNS-1725797) and administered by the Center for Scientific Computing (CSC). The CSC is supported by the California NanoSystems Institute and the Materials Research Science and Engineering Center (MRSEC; NSF DMR 2308708) at UC Santa Barbara. AG, ER, and GF acknowledge the support of project number 101070287 — SWAN-on-chip — HORIZON-CL4-2021-DIGITAL-EMERGING-01; the MUR-PNRR project SAMOTHRACE (ECS00000022), funded by European Union (NextGeneration EU); the projects PRIN 2020LWPKH7 – The Italian factory of micromagnetic modeling and spintronics and PRIN 20225YF2S4 – Magneto-Mechanical Accelerometers, Gyroscopes and Computing based on nanoscale magnetic tunnel junctions (MMAGYC), funded by the Italian Ministry of University and Research. PAL, NAA and MM were supported in part under NSF CCF (grant 1918549). PAL was also supported in part through the  NASA Academic Mission Services (contract NNA16BD14C) under SAA2-403506, as well as the Intelligent Systems Research and Development-3  (ISRDS-3) Contract 80ARC020D0010 under SAA2-403688. JHM acknowledges funding from the VIDI project no. 223.157 (CHASEMAG) and KIC project no. 22016 which are (partly) financed by the Dutch Research Council (NWO), as well as support from the European Union Horizon 2020 and innovation program under the European Research Council ERC Grant Agreement No. 856538 (3D-MAGiC) and the Horizon Europe project no. 101070290 (NIMFEIA). FRT received financial support from the “National Centre for HPC, Big Data and Quantum Computing - HPC”, Project CN\_00000013, CUP B83C22002940006, NRP Mission 4 Component 2 Investment 1.5, funded by the European Union - NextGenerationEU. TS and AR acknowledge support from NSF FuSe2 Award 2425218, Carnegie Mellon University Dean’s Fellowship and Tan Endowed Graduate Fellowship in Electrical and Computer Engineering, Carnegie Mellon University. TS and AR also acknowledge Tong Wu for discussions.

\section{Author Contributions}
SC, MM and KYC conceived the study. KYC supervised the study. SC, AG, ER, NAA performed different parts of the simulations. NAA and SC performed FPGA experiments.  AR and TS performed the physical design simulations for chip design. MMR, FRT, MM analyzed the scaling data with critical feedback. SC wrote the initial draft of the manuscript with inputs from PAL, JHM, MMR, FRT, MC, LST, GF, MM, KYC. All authors contributed to improving the draft and participated in designing the experiments, analyzing the results, and editing the manuscript. 

\section{Competing Interests}
The authors declare no other competing interests.

\balance
\bibliographystyle{unsrtnat}

\begin{thebibliography}{67}
\providecommand{\natexlab}[1]{#1}
\providecommand{\url}[1]{\texttt{#1}}
\expandafter\ifx\csname urlstyle\endcsname\relax
  \providecommand{\doi}[1]{doi: #1}\else
  \providecommand{\doi}{doi: \begingroup \urlstyle{rm}\Url}\fi

\bibitem[Feynman(1982)]{feynman1982simulating}
Richard~P. Feynman.
\newblock {Simulating physics with computers}.
\newblock \emph{International Journal of Theoretical Physics}, 21\penalty0
  (6):\penalty0 467--488, 06 1982.
\newblock ISSN 1572-9575.

\bibitem[Wright et~al.(2022)Wright, Onodera, Stein, Wang, Schachter, Hu, and
  McMahon]{wright2022deep}
Logan~G Wright, Tatsuhiro Onodera, Martin~M Stein, Tianyu Wang, Darren~T
  Schachter, Zoey Hu, and Peter~L McMahon.
\newblock {Deep physical neural networks trained with backpropagation}.
\newblock \emph{Nature}, 601\penalty0 (7894):\penalty0 549--555, 2022.

\bibitem[Aifer et~al.(2024)Aifer, Donatella, Gordon, Duffield, Ahle, Simpson,
  Crooks, and Coles]{aifer2024thermodynamic}
Maxwell Aifer, Kaelan Donatella, Max~Hunter Gordon, Samuel Duffield, Thomas
  Ahle, Daniel Simpson, Gavin Crooks, and Patrick~J Coles.
\newblock {Thermodynamic linear algebra}.
\newblock \emph{npj Unconventional Computing}, 1\penalty0 (1):\penalty0 13,
  2024.

\bibitem[Aadit et~al.(2022{\natexlab{a}})Aadit, Grimaldi, Carpentieri,
  Theogarajan, Martinis, Finocchio, and Camsari]{aadit2022massively}
Navid~Anjum Aadit, Andrea Grimaldi, Mario Carpentieri, Luke Theogarajan, John~M
  Martinis, Giovanni Finocchio, and Kerem~Y Camsari.
\newblock {Massively parallel probabilistic computing with sparse Ising
  machines}.
\newblock \emph{Nature Electronics}, 5\penalty0 (7):\penalty0 460--468,
  2022{\natexlab{a}}.

\bibitem[Mohseni et~al.(2024)Mohseni, Scherer, Johnson, Wertheim, Otten, Aadit,
  Alexeev, Bresniker, Camsari, Chapman, Chatterjee, Dagnew, Esposito, Fahim,
  Fiorentino, Gajjar, Khalid, Kong, Kulchytskyy, Kyoseva, Li, Lott, Markov,
  McDermott, Pedretti, Rao, Rieffel, Silva, Sorebo, Spentzouris, Steiner,
  Torosov, Venturelli, Visser, Webb, Zhan, Cohen, Ronagh, Ho, Beausoleil, and
  Martinis]{Mohseni2024}
Masoud Mohseni, Artur Scherer, K.~Grace Johnson, Oded Wertheim, Matthew Otten,
  Navid~Anjum Aadit, Yuri Alexeev, Kirk~M. Bresniker, Kerem~Y. Camsari, Barbara
  Chapman, Soumitra Chatterjee, Gebremedhin~A. Dagnew, Aniello Esposito, Farah
  Fahim, Marco Fiorentino, Archit Gajjar, Abdullah Khalid, Xiangzhou Kong,
  Bohdan Kulchytskyy, Elica Kyoseva, Ruoyu Li, P.~Aaron Lott, Igor~L. Markov,
  Robert~F. McDermott, Giacomo Pedretti, Pooja Rao, Eleanor Rieffel, Allyson
  Silva, John Sorebo, Panagiotis Spentzouris, Ziv Steiner, Boyan Torosov,
  Davide Venturelli, Robert~J. Visser, Zak Webb, Xin Zhan, Yonatan Cohen, Pooya
  Ronagh, Alan Ho, Raymond~G. Beausoleil, and John~M. Martinis.
\newblock How to build a quantum supercomputer: Scaling from hundreds to
  millions of qubits.
\newblock \emph{arXiv preprint arXiv:2411.10406}, 2024.

\bibitem[Shor(1994)]{shor1994algorithms}
Peter~W Shor.
\newblock {Algorithms for quantum computation: discrete logarithms and
  factoring}.
\newblock In \emph{Proceedings 35th annual symposium on foundations of computer
  science}, pages 124--134. IEEE, 1994.

\bibitem[Arute et~al.(2019)Arute, Arya, Babbush, Bacon, Bardin, Barends,
  Biswas, Boixo, Brandao, Buell, et~al.]{arute2019quantum}
Frank Arute, Kunal Arya, Ryan Babbush, Dave Bacon, Joseph~C Bardin, Rami
  Barends, Rupak Biswas, Sergio Boixo, Fernando~GSL Brandao, David~A Buell,
  et~al.
\newblock {Quantum supremacy using a programmable superconducting processor}.
\newblock \emph{Nature}, 574\penalty0 (7779):\penalty0 505--510, 2019.

\bibitem[Lloyd et~al.(2014)Lloyd, Mohseni, and Rebentrost]{Lloyd_2014}
Seth Lloyd, Masoud Mohseni, and Patrick Rebentrost.
\newblock {Quantum principal component analysis}.
\newblock \emph{Nature Physics}, 10\penalty0 (9):\penalty0 631–633, July
  2014.
\newblock ISSN 1745-2481.

\bibitem[Huang et~al.(2022)Huang, Broughton, Cotler, Chen, Li, Mohseni, Neven,
  Babbush, Kueng, Preskill, and McClean]{Huang_learningexperiments_2022}
Hsin-Yuan Huang, Michael Broughton, Jordan Cotler, Sitan Chen, Jerry Li, Masoud
  Mohseni, Hartmut Neven, Ryan Babbush, Richard Kueng, John Preskill, and
  Jarrod~R. McClean.
\newblock {Quantum advantage in learning from experiments}.
\newblock \emph{Science}, 376\penalty0 (6598):\penalty0 1182--1186, 2022.

\bibitem[Chowdhury et~al.(2023{\natexlab{a}})Chowdhury, Camsari, and
  Datta]{chowdhury2023emulating}
Shuvro Chowdhury, Kerem~Y Camsari, and Supriyo Datta.
\newblock {Emulating Quantum Circuits with Generalized Ising Machines}.
\newblock \emph{IEEE Access}, 2023{\natexlab{a}}.

\bibitem[Troyer and Wiese(2005)]{troyer2005computational}
Matthias Troyer and Uwe-Jens Wiese.
\newblock {Computational Complexity and Fundamental Limitations to Fermionic
  Quantum Monte Carlo Simulations}.
\newblock \emph{Physical review letters}, 94\penalty0 (17):\penalty0 170201,
  2005.

\bibitem[Chowdhury et~al.(2023{\natexlab{b}})Chowdhury, Camsari, and
  Datta]{chowdhury2023accelerated}
Shuvro Chowdhury, Kerem~Y Camsari, and Supriyo Datta.
\newblock {Accelerated quantum Monte Carlo with probabilistic computers}.
\newblock \emph{Communications Physics}, 6\penalty0 (1):\penalty0 85,
  2023{\natexlab{b}}.

\bibitem[Onizawa and Hanyu(2025)]{Onizawa2025}
Naoya Onizawa and Takahiro Hanyu.
\newblock Gpu-accelerated simulated annealing based on p-bits with real-world
  device-variability modeling.
\newblock \emph{Scientific Reports}, 15\penalty0 (1):\penalty0 6118, Feb 2025.
\newblock ISSN 2045-2322.

\bibitem[Block et~al.(2010)Block, Virnau, and Preis]{block2010multi}
Benjamin Block, Peter Virnau, and Tobias Preis.
\newblock {Multi-GPU accelerated multi-spin Monte Carlo simulations of the 2D
  Ising model}.
\newblock \emph{Computer Physics Communications}, 181\penalty0 (9):\penalty0
  1549--1556, 2010.

\bibitem[Preis et~al.(2009)Preis, Virnau, Paul, and Schneider]{preis2009gpu}
Tobias Preis, Peter Virnau, Wolfgang Paul, and Johannes~J Schneider.
\newblock {GPU accelerated Monte Carlo simulation of the 2D and 3D Ising
  model}.
\newblock \emph{Journal of Computational Physics}, 228\penalty0 (12):\penalty0
  4468--4477, 2009.

\bibitem[Yang et~al.(2019)Yang, Chen, Roumpos, Colby, and
  Anderson]{yang2019high}
Kun Yang, Yi-Fan Chen, Georgios Roumpos, Chris Colby, and John Anderson.
\newblock {High performance Monte Carlo simulation of Ising model on TPU
  clusters}.
\newblock In \emph{Proceedings of the International Conference for High
  Performance Computing, Networking, Storage and Analysis}, pages 1--15, 2019.

\bibitem[Fang et~al.(2014)Fang, Feng, Tam, Yun, Moreno, Ramanujam, and
  Jarrell]{fang2014parallel}
Ye~Fang, Sheng Feng, Ka-Ming Tam, Zhifeng Yun, Juana Moreno, Jagannathan
  Ramanujam, and Mark Jarrell.
\newblock {Parallel tempering simulation of the three-dimensional
  Edwards--Anderson model with compact asynchronous multispin coding on GPU}.
\newblock \emph{Computer Physics Communications}, 185\penalty0 (10):\penalty0
  2467--2478, 2014.

\bibitem[Niazi et~al.(2024)Niazi, Chowdhury, Aadit, Mohseni, Qin, and
  Camsari]{niazi2024training}
Shaila Niazi, Shuvro Chowdhury, Navid~Anjum Aadit, Masoud Mohseni, Yao Qin, and
  Kerem~Y. Camsari.
\newblock {Training deep Boltzmann networks with sparse Ising machines}.
\newblock \emph{Nature Electronics}, 7\penalty0 (7):\penalty0 610--619, Jul
  2024.
\newblock ISSN 2520-1131.

\bibitem[Nikhar et~al.(2024)Nikhar, Kannan, Aadit, Chowdhury, and
  Camsari]{nikhar2024all}
Srijan Nikhar, Sidharth Kannan, Navid~Anjum Aadit, Shuvro Chowdhury, and
  Kerem~Y Camsari.
\newblock {All-to-all reconfigurability with sparse and higher-order Ising
  machines}.
\newblock \emph{Nature Communications}, 15\penalty0 (1):\penalty0 8977, 2024.

\bibitem[Aadit et~al.(2023)Aadit, Mohseni, and Camsari]{aadit2023accelerating}
Navid~Anjum Aadit, Masoud Mohseni, and Kerem~Y Camsari.
\newblock {Accelerating Adaptive Parallel Tempering with FPGA-based p-bits}.
\newblock In \emph{2023 IEEE Symposium on VLSI Technology and Circuits (VLSI
  Technology and Circuits)}, pages 1--2. IEEE, 2023.

\bibitem[Aadit et~al.(2021)Aadit, Grimaldi, Carpentieri, Theogarajan,
  Finocchio, and Camsari]{aadit2021computing}
Navid~Anjum Aadit, Andrea Grimaldi, Mario Carpentieri, Luke Theogarajan,
  Giovanni Finocchio, and Kerem~Y Camsari.
\newblock {Computing with invertible logic: Combinatorial optimization with
  probabilistic bits}.
\newblock In \emph{2021 IEEE International Electron Devices Meeting (IEDM)},
  pages 40--3. IEEE, 2021.

\bibitem[Aadit et~al.(2022{\natexlab{b}})Aadit, Grimaldi, Finocchio, and
  Camsari]{aadit2022physics}
Navid~Anjum Aadit, Andrea Grimaldi, Giovanni Finocchio, and Kerem~Y Camsari.
\newblock {Physics-inspired ising computing with ring oscillator activated
  p-bits}.
\newblock In \emph{2022 IEEE 22nd International Conference on Nanotechnology
  (NANO)}, pages 393--396. IEEE, 2022{\natexlab{b}}.

\bibitem[Singh et~al.(2024)Singh, Kobayashi, Cao, Selcuk, Hu, Niazi, Aadit,
  Kanai, Ohno, Fukami, et~al.]{singh2024cmos}
Nihal~Sanjay Singh, Keito Kobayashi, Qixuan Cao, Kemal Selcuk, Tianrui Hu,
  Shaila Niazi, Navid~Anjum Aadit, Shun Kanai, Hideo Ohno, Shunsuke Fukami,
  et~al.
\newblock {CMOS plus stochastic nanomagnets enabling heterogeneous computers
  for probabilistic inference and learning}.
\newblock \emph{Nature Communications}, 15\penalty0 (1):\penalty0 2685, 2024.

\bibitem[Grimaldi et~al.(2022{\natexlab{a}})Grimaldi, Selcuk, Aadit, Kobayashi,
  Cao, Chowdhury, Finocchio, Kanai, Ohno, Fukami,
  et~al.]{grimaldi2022experimental}
Andrea Grimaldi, Kemal Selcuk, Navid~Anjum Aadit, Keito Kobayashi, Qixuan Cao,
  Shuvro Chowdhury, Giovanni Finocchio, Shun Kanai, Hideo Ohno, Shunsuke
  Fukami, et~al.
\newblock {Experimental evaluation of simulated quantum annealing with
  MTJ-augmented p-bits}.
\newblock In \emph{2022 International Electron Devices Meeting (IEDM)}, pages
  22--4. IEEE, 2022{\natexlab{a}}.

\bibitem[Singh et~al.(2023)Singh, Niazi, Chowdhury, Selcuk, Kaneko, Kobayashi,
  Kanai, Ohno, Fukami, and Camsari]{singh2023hardware}
Nihal~Sanjay Singh, Shaila Niazi, Shuvro Chowdhury, Kemal Selcuk, Haruna
  Kaneko, Keito Kobayashi, Shun Kanai, Hideo Ohno, Shunsuke Fukami, and Kerem~Y
  Camsari.
\newblock {Hardware Demonstration of Feedforward Stochastic Neural Networks
  with Fast MTJ-based p-bits}.
\newblock In \emph{2023 International Electron Devices Meeting (IEDM)}, pages
  1--4. IEEE, 2023.

\bibitem[Borders et~al.(2019)Borders, Pervaiz, Fukami, Camsari, Ohno, and
  Datta]{borders2019integer}
William~A Borders, Ahmed~Z Pervaiz, Shunsuke Fukami, Kerem~Y Camsari, Hideo
  Ohno, and Supriyo Datta.
\newblock {Integer factorization using stochastic magnetic tunnel junctions}.
\newblock \emph{Nature}, 573\penalty0 (7774):\penalty0 390--393, 2019.

\bibitem[King et~al.(2021)King, Raymond, Lanting, Isakov, Mohseni,
  Poulin-Lamarre, Ejtemaee, Bernoudy, Ozfidan, Smirnov,
  et~al.]{king2021scaling}
Andrew~D King, Jack Raymond, Trevor Lanting, Sergei~V Isakov, Masoud Mohseni,
  Gabriel Poulin-Lamarre, Sara Ejtemaee, William Bernoudy, Isil Ozfidan,
  Anatoly~Yu Smirnov, et~al.
\newblock {Scaling advantage over path-integral Monte Carlo in quantum
  simulation of geometrically frustrated magnets}.
\newblock \emph{Nature communications}, 12\penalty0 (1):\penalty0 1113, 2021.

\bibitem[Albash and Lidar(2018)]{albash2018demonstration}
Tameem Albash and Daniel~A Lidar.
\newblock {Demonstration of a scaling advantage for a quantum annealer over
  simulated annealing}.
\newblock \emph{Physical Review X}, 8\penalty0 (3):\penalty0 031016, 2018.

\bibitem[Denchev et~al.(2016)Denchev, Boixo, Isakov, Ding, Babbush,
  Smelyanskiy, Martinis, and Neven]{denchev2016computational}
Vasil~S Denchev, Sergio Boixo, Sergei~V Isakov, Nan Ding, Ryan Babbush, Vadim
  Smelyanskiy, John Martinis, and Hartmut Neven.
\newblock What is the computational value of finite-range tunneling?
\newblock \emph{Physical Review X}, 6\penalty0 (3):\penalty0 031015, 2016.

\bibitem[Hen et~al.(2015)Hen, Job, Albash, R{\o}nnow, Troyer, and
  Lidar]{hen2015probing}
Itay Hen, Joshua Job, Tameem Albash, Troels~F R{\o}nnow, Matthias Troyer, and
  Daniel~A Lidar.
\newblock Probing for quantum speedup in spin-glass problems with planted
  solutions.
\newblock \emph{Physical Review A}, 92\penalty0 (4):\penalty0 042325, 2015.

\bibitem[Bernaschi et~al.(2024)Bernaschi, Gonz{\'a}lez-Adalid~Pemart{\'i}n,
  Mart{\'i}n-Mayor, and Parisi]{Bernaschi2024}
Massimo Bernaschi, Isidoro Gonz{\'a}lez-Adalid~Pemart{\'i}n, V{\'i}ctor
  Mart{\'i}n-Mayor, and Giorgio Parisi.
\newblock {The quantum transition of the two-dimensional Ising spin glass}.
\newblock \emph{Nature}, Jul 2024.
\newblock ISSN 1476-4687.

\bibitem[King et~al.(2023)King, Raymond, Lanting, Harris, Zucca, Altomare,
  Berkley, Boothby, Ejtemaee, Enderud, Hoskinson, Huang, Ladizinsky, MacDonald,
  Marsden, Molavi, Oh, Poulin-Lamarre, Reis, Rich, Sato, Tsai, Volkmann,
  Whittaker, Yao, Sandvik, and Amin]{King2023quantum}
Andrew~D. King, Jack Raymond, Trevor Lanting, Richard Harris, Alex Zucca, Fabio
  Altomare, Andrew~J. Berkley, Kelly Boothby, Sara Ejtemaee, Colin Enderud,
  Emile Hoskinson, Shuiyuan Huang, Eric Ladizinsky, Allison J.~R. MacDonald,
  Gaelen Marsden, Reza Molavi, Travis Oh, Gabriel Poulin-Lamarre, Mauricio
  Reis, Chris Rich, Yuki Sato, Nicholas Tsai, Mark Volkmann, Jed~D. Whittaker,
  Jason Yao, Anders~W. Sandvik, and Mohammad~H. Amin.
\newblock {Quantum critical dynamics in a 5,000-qubit programmable spin glass}.
\newblock \emph{Nature}, 617\penalty0 (7959):\penalty0 61--66, April 2023.
\newblock ISSN 1476-4687.

\bibitem[Zhu et~al.(2015{\natexlab{a}})Zhu, Ochoa, and
  Katzgraber]{zhu2015efficient}
Zheng Zhu, Andrew~J Ochoa, and Helmut~G Katzgraber.
\newblock Efficient cluster algorithm for spin glasses in any space dimension.
\newblock \emph{Physical review letters}, 115\penalty0 (7):\penalty0 077201,
  2015{\natexlab{a}}.

\bibitem[Fan et~al.(2023)Fan, Shen, Nussinov, Liu, Sun, and
  Liu]{fan2023searching}
Changjun Fan, Mutian Shen, Zohar Nussinov, Zhong Liu, Yizhou Sun, and Yang-Yu
  Liu.
\newblock Searching for spin glass ground states through deep reinforcement
  learning.
\newblock \emph{Nature communications}, 14\penalty0 (1):\penalty0 725, 2023.

\bibitem[Lee et~al.(2025)Lee, Chowdhury, and Camsari]{Lee2025}
Kyle Lee, Shuvro Chowdhury, and Kerem~Y. Camsari.
\newblock Noise-augmented chaotic ising machines for combinatorial optimization
  and sampling.
\newblock \emph{Communications Physics}, 8\penalty0 (1):\penalty0 35, Jan 2025.
\newblock ISSN 2399-3650.

\bibitem[Houdayer(2001)]{Houdayer2001}
J.~Houdayer.
\newblock {A cluster Monte Carlo algorithm for 2-dimensional spin glasses}.
\newblock \emph{The European Physical Journal B}, 22\penalty0 (4):\penalty0
  479--484, August 2001.
\newblock ISSN 1434-6028.

\bibitem[Zhu et~al.(2015{\natexlab{b}})Zhu, Ochoa, and Katzgraber]{Zhu2015}
Zheng Zhu, Andrew~J. Ochoa, and Helmut~G. Katzgraber.
\newblock {Efficient Cluster Algorithm for Spin Glasses in Any Space
  Dimension}.
\newblock \emph{Physical Review Letters}, 115\penalty0 (7):\penalty0 077201,
  August 2015{\natexlab{b}}.
\newblock ISSN 1079-7114.

\bibitem[King et~al.(2019)King, Mohseni, Bernoudy, Fr{\'e}chette, Sadeghi,
  Isakov, Neven, and Amin]{king2019quantum}
James King, Masoud Mohseni, William Bernoudy, Alexandre Fr{\'e}chette, Hossein
  Sadeghi, Sergei~V Isakov, Hartmut Neven, and Mohammad~H Amin.
\newblock {Quantum-assisted genetic algorithm}.
\newblock \emph{arXiv preprint arXiv:1907.00707}, 2019.

\bibitem[Suzuki(1976)]{suzuki1976relationship}
Masuo Suzuki.
\newblock {Relationship between d-Dimensional Quantal Spin Systems and
  (d+1)-Dimensional Ising Systems Equivalence, Critical Exponents and
  Systematic Approximants of the Partition Function and Spin Correlations}.
\newblock \emph{Progress of Theoretical Physics}, 56:\penalty0 1454--1469,
  1976.

\bibitem[Camsari et~al.(2019)Camsari, Chowdhury, and
  Datta]{camsari2019scalable}
Kerem~Y. Camsari, Shuvro Chowdhury, and Supriyo Datta.
\newblock {Scalable Emulation of Sign-Problem--Free Hamiltonians with
  Room-Temperature $p$-bits}.
\newblock \emph{Phys. Rev. Applied}, 12:\penalty0 034061, 09 2019.

\bibitem[Heim et~al.(2015)Heim, Rønnow, Isakov, and Troyer]{Heim2015quantum}
Bettina Heim, Troels~F. Rønnow, Sergei~V. Isakov, and Matthias Troyer.
\newblock {Quantum versus classical annealing of Ising spin glasses}.
\newblock \emph{Science}, 348\penalty0 (6231):\penalty0 215--217, 2015.

\bibitem[Santoro et~al.(2002)Santoro, Martoňák, Tosatti, and
  Car]{Santoro2002}
Giuseppe~E. Santoro, Roman Martoňák, Erio Tosatti, and Roberto Car.
\newblock {Theory of Quantum Annealing of an Ising Spin Glass}.
\newblock \emph{Science}, 295\penalty0 (5564):\penalty0 2427--2430, 2002.

\bibitem[Billoire et~al.(2018)Billoire, Fernandez, Maiorano, Marinari,
  Martin-Mayor, Moreno-Gordo, Parisi, Ricci-Tersenghi, and
  Ruiz-Lorenzo]{Billoire_2018}
A~Billoire, L~A Fernandez, A~Maiorano, E~Marinari, V~Martin-Mayor,
  J~Moreno-Gordo, G~Parisi, F~Ricci-Tersenghi, and J~J Ruiz-Lorenzo.
\newblock {Dynamic variational study of chaos: spin glasses in three
  dimensions}.
\newblock \emph{Journal of Statistical Mechanics: Theory and Experiment},
  2018\penalty0 (3):\penalty0 033302, mar 2018.

\bibitem[Papakonstantinou and Malakis(2014)]{Papakonstantinou_2014}
T~Papakonstantinou and A~Malakis.
\newblock {Parallel tempering and 3D spin glass models}.
\newblock \emph{Journal of Physics: Conference Series}, 487\penalty0
  (1):\penalty0 012010, mar 2014.

\bibitem[Earl and Deem(2005)]{earl2005parallel}
David~J Earl and Michael~W Deem.
\newblock {Parallel tempering: Theory, applications, and new perspectives}.
\newblock \emph{Physical Chemistry Chemical Physics}, 7\penalty0 (23):\penalty0
  3910--3916, 2005.

\bibitem[Swendsen and Wang(1986)]{swendsen1986replica}
Robert~H Swendsen and Jian-Sheng Wang.
\newblock {Replica Monte Carlo simulation of spin-glasses}.
\newblock \emph{Physical review letters}, 57\penalty0 (21):\penalty0 2607,
  1986.

\bibitem[Hukushima and Nemoto(1996)]{hukushima1996exchange}
Koji Hukushima and Koji Nemoto.
\newblock {Exchange Monte Carlo method and application to spin glass
  simulations}.
\newblock \emph{Journal of the Physical Society of Japan}, 65\penalty0
  (6):\penalty0 1604--1608, 1996.

\bibitem[Grimaldi et~al.(2022{\natexlab{b}})Grimaldi, S\'anchez-Tejerina,
  Anjum~Aadit, Chiappini, Carpentieri, Camsari, and
  Finocchio]{andrea2022spintronics}
Andrea Grimaldi, Luis S\'anchez-Tejerina, Navid Anjum~Aadit, Stefano Chiappini,
  Mario Carpentieri, Kerem Camsari, and Giovanni Finocchio.
\newblock Spintronics-compatible approach to solving maximum-satisfiability
  problems with probabilistic computing, invertible logic, and parallel
  tempering.
\newblock \emph{Phys. Rev. Appl.}, 17:\penalty0 024052, Feb 2022{\natexlab{b}}.

\bibitem[Isakov et~al.(2015)Isakov, Zintchenko, Rønnow, and
  Troyer]{Isakov2015optimising}
S.V. Isakov, I.N. Zintchenko, T.F. Rønnow, and M.~Troyer.
\newblock {Optimised simulated annealing for Ising spin glasses}.
\newblock \emph{Computer Physics Communications}, 192:\penalty0 265--271, 2015.
\newblock ISSN 0010-4655.

\bibitem[Katzgraber et~al.(2006)Katzgraber, Trebst, Huse, and
  Troyer]{Katzgraber_2006}
Helmut~G Katzgraber, Simon Trebst, David~A Huse, and Matthias Troyer.
\newblock Feedback-optimized parallel tempering monte carlo.
\newblock \emph{Journal of Statistical Mechanics: Theory and Experiment},
  2006\penalty0 (03):\penalty0 P03018, 2006.

\bibitem[Mohseni et~al.(2021)Mohseni, Eppens, Strumpfer, Marino, Denchev, Ho,
  Isakov, Boixo, Ricci-Tersenghi, and Neven]{Mohseni2021}
Masoud Mohseni, Daniel Eppens, Johan Strumpfer, Raffaele Marino, Vasil Denchev,
  Alan~K. Ho, Sergei~V. Isakov, Sergio Boixo, Federico Ricci-Tersenghi, and
  Hartmut Neven.
\newblock {Nonequilibrium Monte Carlo for Unfreezing Variables in Hard
  Combinatorial Optimization}.
\newblock \emph{arXiv}, nov 2021.

\bibitem[Hayakawa et~al.(2021)Hayakawa, Kanai, Funatsu, Igarashi, Jinnai,
  Borders, Ohno, and Fukami]{hayakawa2021nanosecond}
Keisuke Hayakawa, Shun Kanai, Takuya Funatsu, Junta Igarashi, Butsurin Jinnai,
  WA~Borders, H~Ohno, and S~Fukami.
\newblock {Nanosecond random telegraph noise in in-plane magnetic tunnel
  junctions}.
\newblock \emph{Physical review letters}, 126\penalty0 (11):\penalty0 117202,
  2021.

\bibitem[Soumah et~al.(2024)Soumah, Desplat, Phan, Valli, Madhavan, Disdier,
  Auffret, Sousa, Ebels, and Talatchian]{soumah2024nanosecond}
Lucile Soumah, Louise Desplat, Nhat-Tan Phan, Ahmed Sidi~El Valli, Advait
  Madhavan, Florian Disdier, St{\'e}phane Auffret, Ricardo Sousa, Ursula Ebels,
  and Philippe Talatchian.
\newblock {Entropy-Assisted Nanosecond stochastic operation in perpendicular
  superparamagnetic tunnel junctions}.
\newblock \emph{arXiv preprint arXiv:2402.03452}, 2024.

\bibitem[Hamze et~al.(2018)Hamze, Jacob, Ochoa, Perera, Wang, and
  Katzgraber]{hamze2018near}
Firas Hamze, Darryl~C Jacob, Andrew~J Ochoa, Dilina Perera, Wenlong Wang, and
  Helmut~G Katzgraber.
\newblock From near to eternity: spin-glass planting, tiling puzzles, and
  constraint-satisfaction problems.
\newblock \emph{Physical Review E}, 97\penalty0 (4):\penalty0 043303, 2018.

\bibitem[Baity-Jesi et~al.(2014)Baity-Jesi, Baños, Cruz, Fernandez,
  Gil-Narvion, Gordillo-Guerrero, Iñiguez, Maiorano, Mantovani, Marinari,
  Martin-Mayor, Monforte-Garcia, {Muñoz Sudupe}, Navarro, Parisi,
  Perez-Gaviro, Pivanti, Ricci-Tersenghi, Ruiz-Lorenzo, Schifano, Seoane,
  Tarancon, Tripiccione, and Yllanes]{BAITYJESI2014550}
M.~Baity-Jesi, R.A. Baños, A.~Cruz, L.A. Fernandez, J.M. Gil-Narvion,
  A.~Gordillo-Guerrero, D.~Iñiguez, A.~Maiorano, F.~Mantovani, E.~Marinari,
  V.~Martin-Mayor, J.~Monforte-Garcia, A.~{Muñoz Sudupe}, D.~Navarro,
  G.~Parisi, S.~Perez-Gaviro, M.~Pivanti, F.~Ricci-Tersenghi, J.J.
  Ruiz-Lorenzo, S.F. Schifano, B.~Seoane, A.~Tarancon, R.~Tripiccione, and
  D.~Yllanes.
\newblock {Janus II: A new generation application-driven computer for
  spin-system simulations}.
\newblock \emph{Computer Physics Communications}, 185\penalty0 (2):\penalty0
  550--559, 2014.
\newblock ISSN 0010-4655.

\bibitem[Sutton et~al.(2020)Sutton, Faria, Ghantasala, Jaiswal, Camsari, and
  Datta]{sutton2020autonomous}
Brian Sutton, Rafatul Faria, Lakshmi~Anirudh Ghantasala, Risi Jaiswal,
  Kerem~Yunus Camsari, and Supriyo Datta.
\newblock {Autonomous probabilistic coprocessing with petaflips per second}.
\newblock \emph{IEEE Access}, 8:\penalty0 157238--157252, 2020.

\bibitem[Iyer and Achour(2025)]{iyer2025efficient}
Devrath Iyer and Sara Achour.
\newblock Efficient optimization with encoded ising models.
\newblock In \emph{2025 IEEE International Symposium on High Performance
  Computer Architecture (HPCA)}, pages 85--98. IEEE, 2025.

\bibitem[Clark et~al.(2016)Clark, Vashishtha, Shifren, Gujja, Sinha, Cline,
  Ramamurthy, and Yeric]{asap7}
Lawrence~T. Clark, Vinay Vashishtha, Lucian Shifren, Aditya Gujja, Saurabh
  Sinha, Brian Cline, Chandarasekaran Ramamurthy, and Greg Yeric.
\newblock {ASAP7: A 7-nm finFET predictive process design kit}.
\newblock \emph{Microelectronics Journal}, 53:\penalty0 105--115, 2016.
\newblock ISSN 1879-2391.

\bibitem[Camsari et~al.(2017)Camsari, Faria, Sutton, and
  Datta]{camsari2017stochastic}
Kerem~Yunus Camsari, Rafatul Faria, Brian~M Sutton, and Supriyo Datta.
\newblock {Stochastic p-bits for invertible logic}.
\newblock \emph{Physical Review X}, 7\penalty0 (3):\penalty0 031014, 2017.

\bibitem[Br{\'{e}}laz(1979)]{Brelaz1979}
Daniel Br{\'{e}}laz.
\newblock {New methods to color the vertices of a graph}.
\newblock \emph{Communications of the {ACM}}, 22\penalty0 (4):\penalty0
  251--256, apr 1979.

\bibitem[Melchert(2009)]{melchert2009}
O.~Melchert.
\newblock autoscale.py - a program for automatic finite-size scaling analyses:
  A user's guide.
\newblock \emph{arXiv preprint arXiv:0910.5403}, 2009.

\bibitem[Rieger and Kawashima(1999)]{Rieger1999}
H.~Rieger and N.~Kawashima.
\newblock {Application of a continuous time cluster algorithm to the
  two-dimensional random quantum Ising ferromagnet}.
\newblock \emph{The European Physical Journal B - Condensed Matter and Complex
  Systems}, 9\penalty0 (2):\penalty0 233--236, May 1999.

\bibitem[Chowdhury et~al.(2023{\natexlab{c}})Chowdhury, Grimaldi, Aadit, Niazi,
  Mohseni, Kanai, Ohno, Fukami, Theogarajan, Finocchio, Datta, and
  Camsari]{chowdhury2023fullstack}
Shuvro Chowdhury, Andrea Grimaldi, Navid~Anjum Aadit, Shaila Niazi, Masoud
  Mohseni, Shun Kanai, Hideo Ohno, Shunsuke Fukami, Luke Theogarajan, Giovanni
  Finocchio, Supriyo Datta, and Kerem~Y. Camsari.
\newblock {A Full-Stack View of Probabilistic Computing With p-Bits: Devices,
  Architectures, and Algorithms}.
\newblock \emph{IEEE Journal on Exploratory Solid-State Computational Devices
  and Circuits}, 9\penalty0 (1):\penalty0 1--11, 2023{\natexlab{c}}.

\bibitem[Sajeeb et~al.(2025)Sajeeb, Aadit, Chowdhury, Wu, Smith, Chinmay, Raut,
  Camsari, Delacour, and Srimani]{sajeeb2025scalable}
M.~Mahmudul~Hasan Sajeeb, Navid~Anjum Aadit, Shuvro Chowdhury, Tong Wu, Cesely
  Smith, Dhruv Chinmay, Atharva Raut, Kerem~Y. Camsari, Corentin Delacour, and
  Tathagata Srimani.
\newblock Scalable connectivity for ising machines: Dense to sparse.
\newblock \emph{Phys. Rev. Appl.}, 24:\penalty0 014005, Jul 2025.

\bibitem[Rønnow et~al.(2014)Rønnow, Wang, Job, Boixo, Isakov, Wecker,
  Martinis, Lidar, and Troyer]{ronnow_defining_2014}
Troels~F. Rønnow, Zhihui Wang, Joshua Job, Sergio Boixo, Sergei~V. Isakov,
  David Wecker, John~M. Martinis, Daniel~A. Lidar, and Matthias Troyer.
\newblock {Defining and detecting quantum speedup}.
\newblock \emph{Science}, 345\penalty0 (6195):\penalty0 420--424, July 2014.

\bibitem[King et~al.(2022)King, Suzuki, Raymond, Zucca, Lanting, Altomare,
  Berkley, Ejtemaee, Hoskinson, Huang, Ladizinsky, MacDonald, Marsden, Oh,
  Poulin-Lamarre, Reis, Rich, Sato, Whittaker, Yao, Harris, Lidar, Nishimori,
  and Amin]{king_coherent_2022}
Andrew~D. King, Sei Suzuki, Jack Raymond, Alex Zucca, Trevor Lanting, Fabio
  Altomare, Andrew~J. Berkley, Sara Ejtemaee, Emile Hoskinson, Shuiyuan Huang,
  Eric Ladizinsky, Allison J.~R. MacDonald, Gaelen Marsden, Travis Oh, Gabriel
  Poulin-Lamarre, Mauricio Reis, Chris Rich, Yuki Sato, Jed~D. Whittaker, Jason
  Yao, Richard Harris, Daniel~A. Lidar, Hidetoshi Nishimori, and Mohammad~H.
  Amin.
\newblock {Coherent quantum annealing in a programmable 2,000 qubit {Ising}
  chain}.
\newblock \emph{Nature Physics}, 18\penalty0 (11):\penalty0 1324--1328,
  November 2022.
\newblock ISSN 1745-2481.
\newblock Number: 11 Publisher: Nature Publishing Group.

\bibitem[Carsello et~al.(2022)Carsello, Thomas, Nayak, Chen, Horowitz, Raina,
  and Torng]{mflowgen}
Alex Carsello, James Thomas, Ankita Nayak, Po-Han Chen, Mark Horowitz, Priyanka
  Raina, and Christopher Torng.
\newblock {mflowgen: a modular flow generator and ecosystem for
  community-driven physical design: invited}.
\newblock In \emph{Proceedings of the 59th ACM/IEEE Design Automation
  Conference}, DAC '22, page 1339–1342, New York, NY, USA, 2022. Association
  for Computing Machinery.
\newblock ISBN 9781450391429.

\end{thebibliography}

\clearpage
\pagebreak
\newpage

\setcounter{secnumdepth}{3}

\newcommand{\beginsupplement}{%
        \setcounter{table}{0}
        \renewcommand{\thetable}{S\arabic{table}}%
        \setcounter{figure}{0}
        \renewcommand{\thefigure}{S\arabic{figure}}%
        \setcounter{equation}{0}
        \renewcommand{\theequation}{S.\arabic{equation}}%
        \renewcommand{\thealgocf}{S\arabic{algocf}}
        \setcounter{algocf}{0}%
     }

\onecolumngrid
\begin{center}
{\sffamily\Large\bf Supplementary Information\par}
\vskip 0.5em
{\sffamily\LARGE\bf Pushing the Boundary of Quantum Advantage in Hard Combinatorial Optimization with Probabilistic Computers \par}
\vspace{1em}
\normalfont\noindent{\sffamily Shuvro Chowdhury, Navid Anjum Aadit, Andrea Grimaldi, Eleonora Raimondo, Atharva Raut, P. Aaron Lott, Johan H. Mentink,  Marek M. Rams, Federico Ricci-Tersenghi, Massimo Chiappini, Luke S. Theogarajan, Tathagata Srimani, Giovanni Finocchio, Masoud Mohseni and Kerem Y. Camsari}
\end{center}

\beginsupplement
\renewcommand{\theHfigure}{S\arabic{figure}}
\renewcommand{\theHtable}{S\arabic{table}}
\renewcommand{\theHequation}{S.\arabic{equation}}
\renewcommand{\theHalgocf}{S\arabic{algocf}}
\renewcommand{\theHsection}{S\arabic{figure}}

\setcounter{section}{0}
\setcounter{subsection}{0}

\vspace{-15pt}

\section{Discrete-time Simulated Quantum Annealing (DT-SQA) algorithm}
\label{supp_sec:DTPIMC_algo}
The DT-SQA algorithm used in Fig.~\ref{fig:scaling_SQA} of the main text relies on the Suzuki-Trotter approximation, where the partition function, $Z_\text{Q}$ of a quantum Hamiltonian, 
\begin{align} 
    H_{\text{Q}} = -\sum\limits_{i<j}{J_{ij}\sigma_i^z\sigma^z_j}-\Gamma_x\sum\limits_{i}\sigma^x_i
\end{align} with $\sigma_i^{\alpha}$ ($\alpha \in \{x,y,z\}$) being Pauli spin matrix at site $i$, is approximated by the partition function, $Z_{\text{C}}$ of a classical Hamiltonian, $H_\text{C}$ as
\begin{align}
Z_\text{Q}= \text{tr}\left[\exp{(-\beta H_{\text{Q}})}\right] = \lim_{R\to\infty}Z_\text{C}= \lim_{R\to\infty}\exp{(-\beta H_{\text{C}})}
\end{align}
where $H_\text{C}$, corresponding to $H_\text{Q}$, is defined as 
\begin{align}
H_{\text{C}} = -\sum\limits_{k=1}^{R}\left[\sum\limits_{i < j}{J}_{\parallel ,ij}{\sigma}_{i,k}{\sigma}_{j,k}+\sum\limits_{i}{J}_{\perp }{\sigma}_{i,k}{\sigma}_{i,k+1}\right] \text{ with } J_{\parallel,ij}=\cfrac{J_{ij}}{R} \text{ and } J_{\perp}=-\cfrac{1}{2\beta}\ln{\left[\tanh{\left(\cfrac{\beta\Gamma_x}{R}\right)}\right]}. 
\label{supp_eq:SQA_ham}
\end{align}
where $\sigma_{i,j}$ (without any superscript) denotes Ising spin ($\sigma_{i,j}\in\{-1,+1\}$) at site $i$ of replica $j$.  Thus, a $d$-dimensional quantum Hamiltonian can be theoretically mapped to a ($d+1$)-dimensional classical Hamiltonian, where the extra dimension corresponds to the replica dimension. When $R$ approaches infinity, the partition function corresponding to the quantum Hamiltonian is exactly reproduced by the partition function of the mapped classical counterpart. However, for practical purposes or when DT-SQA is used as a classical algorithm (for example, in optimization), a finite number of replicas typically in the range of 10 to 100--may be preferred \cite{camsari2019scalable,Santoro2002, Heim2015quantum}. A pseudocode outlining our implementation of the DT-SQA algorithm is presented in Algorithm~\ref{algo:DTPIMC}. As described in the Methods section, graph coloring is used to partition the spin system into independent sets that can be updated in parallel. Each color corresponds to one of these independent sets. Algorithm~\ref{algo:DTPIMC} implements this update scheme by looping over colors.

There is also a continuous-time version of the DT-SQA algorithm (CT-SQA) \cite{Rieger1999}. However, CT-SQA lacks the straightforward hardware implementation offered by DT-SQA. Although both algorithms emulate the equilibrium statistics of the quantum system, differences arise when comparing their transient dynamics to those of a quantum annealer. 
\SetKwComment{Comment}{\normalfont $\triangleright$ }{}
\SetKwInput{KwInput}{Input}                
\SetKwInput{KwOutput}{Output}              
\SetKwFor{ParFor}{for}{do in parallel}{end}
\begin{algorithm}[!b]
    \DontPrintSemicolon
    \KwInput{Weights, biases, number of replicas, number of sweeps, colormap, annealing schedule, temperature}
    
    \KwOutput{State corresponding to minimum energy, ${m}_{\mathrm{opt}}$}
    \SetKwFunction{FMain}{p-computer}
    \SetKwProg{Fn}{Function}{:}{}
    \Fn{\FMain{weights, biases, colormap, temp.}}{
        \For{each color in the colormap}{
            \For{each p-bit in the color}{ 
                solve Eq.~(\ref{eq:synapse}) and Eq.~(\ref{eq:neuron}).\;
            }
        }
    }

    Divide weights by number of replicas.\;
    Generate all replicas as indicated in number of replicas.\;
    Insert transverse couplings between the replicas.\;
    Perform graph coloring for the replicated network given the colormap of a single replica.\;
    Initialize all spins in all replicas to random states.\;
    \For{each sweep until number of sweeps is reached}{
        Get the transverse field value from the annealing schedule.\;
        Update the transverse coupling using the transverse field.\;
        Sample p-bit states from p-computer.\;     
    }
    Compute energy of each replica.\;
    \KwRet p-bit states for the replica with the minimum energy.\;
\caption{Discrete-time simulated quantum annealing algorithm with p-computers}
\label{algo:DTPIMC}
\end{algorithm}

\subsection{Residual energy as a function of total Monte Carlo effort}

In the main text, we analyze the scaling behavior of the residual energy as a function of Monte Carlo sweeps for three distinct Discrete-Time Simulated Quantum Annealing (DT-SQA) systems, each characterized by a different number of Trotter slices. We define a Monte Carlo sweep as the process of updating every spin in the system exactly once. Consequently, larger systems naturally involve more computational effort, as they necessitate a greater number of probabilistic spin flips per sweep.

This definition aligns intuitively with our hardware-centric context. We envision the different systems, each comprising $R$ replicas, as separate black-box computational units or stand-alone chips. As previously detailed in Refs.~\cite{aadit2022massively,nikhar2024all,chowdhury2023fullstack, sajeeb2025scalable}, our graph-colored architecture is designed to mimic a truly asynchronous analog probabilistic annealer (one example being a stochastic Magnetic Tunnel Junction based probabilistic computer \cite{borders2019integer}) and as such it enables updates across the entire network in constant time. 

This uniform update time primarily depends on the topology of the network rather than system size. Specifically, if a given graph has degree scaling as $\mathcal{O}(k)$, where $k$ denotes the number of neighbors per node (as in the case of 3D spin glasses considered in this work), the clock frequency that sets the update frequency is mainly determined by the time required to compute the local field as given by Eq.~(\ref{eq:synapse}), a calculation independent of the overall network size, as we demonstrate in Supplementary Fig.~\ref{fig:chip_area} and Supplementary Table~\ref{tab:mflowgen_results}. 

Our use of Monte Carlo sweeps for systems of different sizes is different from how Monte Carlo effort is defined in CPU-centric comparisons \cite{ronnow_defining_2014}. As carefully discussed in Ref.~\cite{ronnow_defining_2014}, an analog quantum annealer acquires an $\mathcal{O}(N)$ speedup (similar to our asynchronous architecture) due to its linearly growing resources with problem size and this parallel speedup must be separated from intrinsically quantum speedups. Given that our systems also benefit from the same speedup, our definitional choice for MCS is appropriate. 

\begin{figure}[!t]
    \centering
    \vspace{0pt}
    \includegraphics[width=3.5in,keepaspectratio]{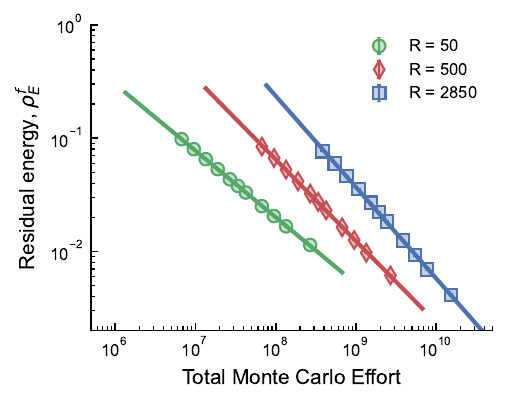}
    \caption{\footnotesize\textbf{Slope of the residual energy, $\rho_E^f$ as a function of total Monte Carlo effort:} The residual energy is re-plotted against the total Monte Carlo effort for the logical instances of size $15\times15\times12$. Here total Monte Carlo effort is defined as the total number of attempted spin flips during the whole annealing process. We emphasize that doing so only changes the prefactor  and does not change the slope of the plots because of the power-law nature since multiplying time with any constant factor does not change the slope. We also emphasize that the Total Monte Carlo effort is a fixed-size CPU-centric measure, as discussed in Ref.~\cite{ronnow_defining_2014}, not appropriate in our context, as we discuss in the text.}
    \label{fig:SQA_logical_with_R}
\end{figure}

Nevertheless, it is crucial to note that our scaling argument against quantum annealing does not hinge upon any MCS definition, since the residual energy comparisons are made on a fixed size, hence $N$ does not vary. As demonstrated in Fig.~\ref{fig:SQA_logical_with_R}, the residual energy exhibits a clear power-law dependence on annealing time or Monte Carlo sweeps. Thus, multiplying the annealing duration by a constant factor (in this case, the number of replicas $R$) does not alter the fundamental slope of the scaling:

\begin{align}
\rho_\mathrm{E}^\mathrm{f}(Rt_\mathrm{a}) \propto (Rt_\mathrm{a})^{-\kappa_\mathrm{f}} = \mathrm{(constant)} \ t_\mathrm{a}^{-\kappa_\mathrm{f}}.
\end{align}

\subsection{Performance of DT-SQA algorithm on embedded instances}
\begin{figure*}[!ht]
    \centering
    \vspace{0pt}
    \includegraphics[width=0.95\textwidth,keepaspectratio]{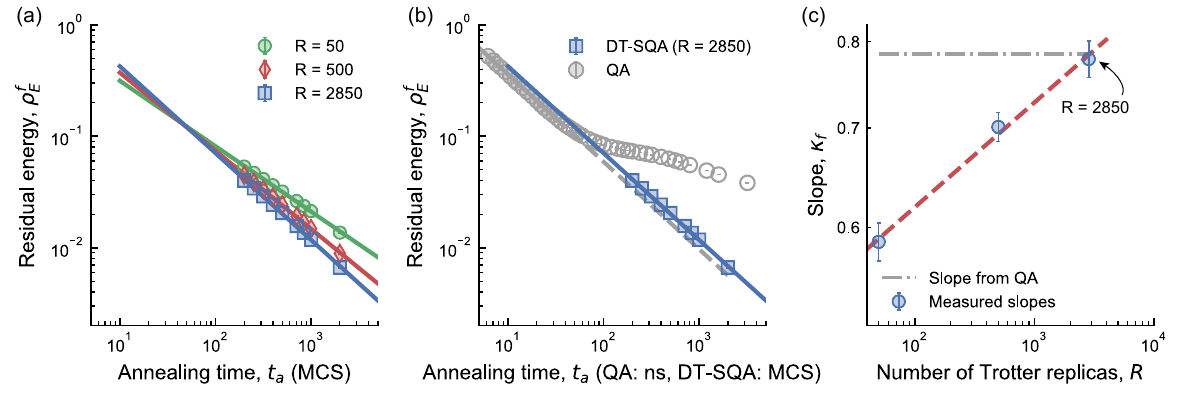}
    \caption{\footnotesize\textbf{Replica scaling of the DT-SQA on embedded instances:} (a) Residual energy, $\rho_{\mathrm{E}}^{\mathrm{f}}$, plotted as a function of the annealing time, $t_a$, for embedded instances of size $15\times15\times12\times2$ (the latter 2 represents the number of physical qubits used to represent a logical spin), for three distinct values of  $R$. In these instances, each lattice point consists of a pair of spins strongly coupled by a ferromagnetic interaction with an absolute strength of 2. (b) The DT-SQA curve with $R=2850$ in (a) is compared to the slope obtained from a quantum annealer, showing nearly identical slope as reported in Ref.~\cite{King2023quantum}. (c) The measured slopes ($\kappa_{\mathrm{f}}$) from DT-SQA simulations are plotted against the number of replicas, $R$. The red dashed line serves as a visual guide, while the gray dotted-dashed line represents the slope derived from the QA. The slope for the quantum annealer can be matched by using more than 2850 replicas.}
    \label{fig:SQA_embedded2}
\end{figure*}

In Fig.~\ref{fig:scaling_SQA} of the main text, we presented scaling results for logical instances using the DT-SQA algorithm. Fig.~\ref{fig:SQA_embedded2} shows the scaling results from DT-SQA experiments on embedded instances. As in logical instances, the slope increases gradually with the number of replicas for the embedded instances, however, achieving the same slope as the quantum annealer requires more replicas compared to logical instances. Also, the embedded instances as provided require four colors for graph coloring (due to the embedding requirements). An even number of replicas are used so that the replicated networks of the embedded instances can also be colored with four colors.

\subsection{Cube size and Trotter replica dependence of DT-SQA}
\begin{figure*}[!ht]
    \centering
    \vspace{0pt}
    \includegraphics[width=0.95\textwidth,keepaspectratio]{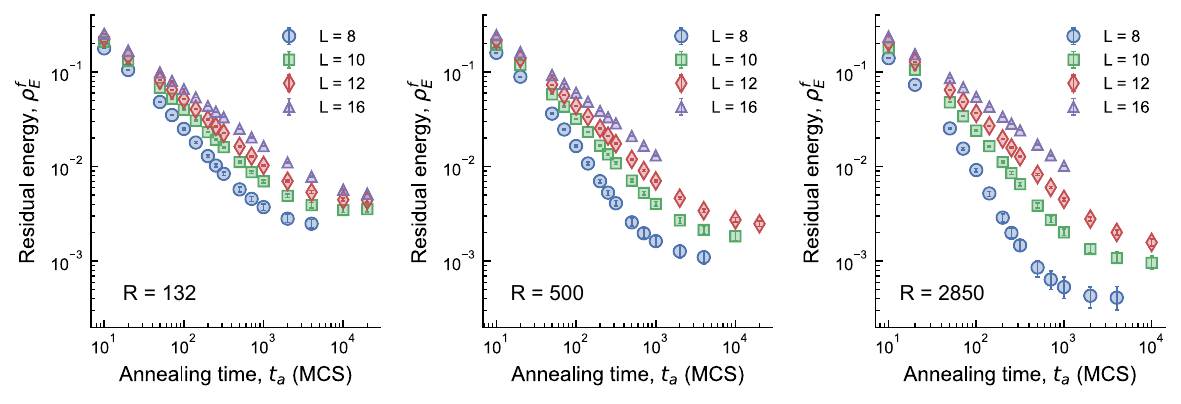}    \caption{\footnotesize\textbf{Cube size and  replica dependence of DT-SQA:} Residual energy ($\rho_\mathrm{E}^\mathrm{f}$) is plotted as a function of annealing time ($t_\mathrm{a}$) for various cube sizes ($L$) and three different number of Trotter replicas ($R$). Each data point represents an average over 300 problem instances, with each instance averaged over 50 independent runs.  Error bars represent the 95\% bootstrap confidence interval of the mean across 300 spin-glass instances.}
    \label{fig:DTSQA_L_R}
\end{figure*}

Fig.~\ref{fig:DTSQA_L_R} illustrates the dependence of the final residual energy on annealing time ($t_a$) for different cube sizes ($L$) of logical instances. The slopes (absolute value) of the plots decrease as $L$ increases, but all plots eventually reach a flat plateau region. This behavior is consistently observed across different numbers of Trotter replicas ($R$) as shown in the figure. The residual energy at which the plateau occurs decreases with an increasing number of Trotter replicas, also observed in \cite{Heim2015quantum}.

\section{Analysis of DT-SQA scaling using extreme value theory}
\label{sec:supp_evt}

\begin{figure*}[!ht]
    \centering
    \vspace{0pt}
    \includegraphics[width=5in,keepaspectratio]{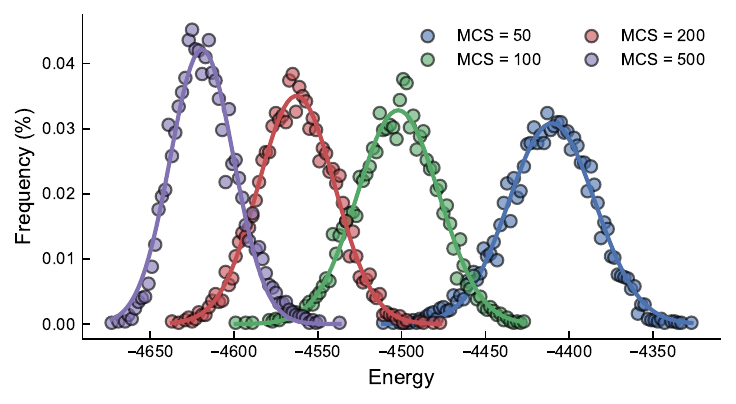}
    \caption{\footnotesize\textbf{Gaussian distribution of minimum energy values:} The distribution of minimum energies for a randomly chosen spin-glass instance is shown at various MCS values, based on 5000 independent runs. In each run, DT-SQA is performed with 32 Trotter replicas for annealing at a fixed MCS, and the best energy replica is selected at the end of the annealing process. The distributions closely follow a Gaussian shape for smaller MCS values, as indicated by the dashed lines representing Gaussian fits. At higher MCS values, the distributions become increasingly skewed to the left due to the hard constraint imposed by the ground state energy. }
    \label{fig:gaussian_EVT}
\end{figure*}

In this section, we explain the increase in DT-SQA slopes with increasing Trotter replicas using extreme value theory (EVT). Recall that in our DT-SQA simulations, we have $R$ interconnected Trotter replicas and we select the replica with the best (minimum) energy. The interconnection of Trotter replicas complicates the direct application of EVT. Therefore, we first describe the conventional EVT, followed by the modified EVT for interconnected Trotter replicas. 

\textbf{Conventional EVT:} A straightforward application of EVT can be demonstrated with the following experiment: we run $P$ independent copies of DT-SQA simulations, each consisting of $R$ Trotter replicas. From each of these $P$ independent copies, we select the minimum energy replica and then choose the minimum among these $P$ minimum energies. Since these $P$ minimum energies are independent and identically distributed, their distribution is approximately a Gaussian  distribution, particularly at low MCS as shown in Supplementary Fig.~\ref{fig:gaussian_EVT}. At high MCS, the distributions become skewed to the left, due to the hard constraint imposed by the ground state which serves as a lower bound. 

For the derivations that follow, we assume a Gaussian distribution for the energies. Let $E_1, E_2,\ldots,E_{P}$ represent the energies of $P$ independent and identically distributed (i.i.d.) Gaussian random variables, each characterized by a mean $\mu$ and standard deviation $\sigma$. Our objective is to derive the expected value of the minimum of these $P$ runs, $\mathbb{E}\left(\text{min}\left(E_1, E_2,\ldots,E_{P}\right)\right)$ which plays a crucial role in improving the residual energy and, consequently the slope in DT-SQA simulations. The probability density function (PDF) of a Gaussian random variable $E_i$ is given by:
\begin{align}
f\left(x\right) = \cfrac{1}{\sqrt{2\pi}\sigma}\, \exp{\left(-\frac{\left(x-\mu\right)^2}{2\sigma^2}\right)}
\end{align}

The cumulative distribution function (CDF) of a random variable $E$ represents the probability that $E$ takes a value less than or equal to a specific value $x$. Mathematically, the CDF, $F(x)$, is defined as:
\begin{align}
F(x) = \text{Pr}\left(E\leq x\right)
\end{align}

For a continuous random variable, the CDF can be obtained by integrating its probability density function (PDF) $f(x)$:
\begin{align}
F(x) = \int_{-\infty}^{x}{f(x)\,\mathrm{d}x}
\end{align} 
The CDF satisfies the property that 
\begin{align}
\lim_{x\to\infty}F\left(x\right) = 1
\label{eq:CDF_prop}
\end{align} 
which ensures that the total probability adds to 1. For a Gaussian random variable, the cumulative distribution function (CDF) is given by:
\begin{align}
F\left(x\right) = \cfrac{1}{\sqrt{2\pi}\sigma}\,\int_{-\infty}^{x}{\exp{\left(-\frac{\left(x-\mu\right)^2}{2\sigma^2}\right)}\,\mathrm{d}x} = \cfrac{1}{2}\,\left[1+\mathrm{erf}{\left(\cfrac{x-\mu}{\sqrt{2}\sigma}\right)}\right]
\end{align}
where $\mathrm{erf}(\cdot)$ is the error function.

The CDF allows us to compute probabilities over intervals and is particularly useful to find the probability that the minimum of $P$ variables falls below a given threshold. To calculate the probability that the minimum of these $P$ random variables $E_i$ is less than or equal to a certain threshold $x$, i.e., $\text{Pr}(\text{min}\left(E_1,E_2,\ldots,E_P\right)\leq x)$, we start by noting that for a single random variable $E_i$, $\text{Pr}(\text{min}\left(E_i\right)\leq x) = \text{Pr}(E_i\leq x)$ and is given by its CDF. For two variables, $E_1$ and $E_2$, the probability that their minimum is less than or equal to $x$ can be computed using their complement probabilities, similar to the Bernoulli case. Specifically, 
\begin{align}
\text{Pr}(\text{min}\left(E_1,E_2\right)\leq x) = 1- \text{Pr}(\text{min}\left(E_1,E_2\right)> x)
\end{align}
But $\text{min}\left(E_1,E_2\right)> x$ implies that both $E_1, E_2 >x$ and therefore,
\begin{align}
\text{Pr}(\text{min}\left(E_1,E_2\right)\leq x) = 1-\left[\text{Pr}\,(E_1 > x)\,\text{Pr}\,(E_2 > x)\right]
\end{align}

Using the property  $\text{Pr}\,(E_i > x) = 1-\text{Pr}\,(E_i \leq x) = 1-F(x)$, this can be rewritten as:
\begin{align}
\text{Pr}\,(\text{min}\left(E_1,E_2\right)\leq x) = 1-\left\{\left[1-\text{Pr}\,(E_1 \leq x)\right]\left[1-\text{Pr}\,(E_2 \leq x)\right]\right\} = 1-[1-F(x)]^2
\end{align}

Generalizing this result to $P$ independent random variables, we obtain:
\begin{align}
\text{Pr}\,(\text{min}\left(E_1,E_2,\ldots,E_P\right)\leq x) = F_P\left(x\right) = 1-\left[1-F\left(x\right)\right]^P=1-2^{-P}\left[\mathrm{erfc}\left(\cfrac{x-\mu}{\sqrt{2}\sigma}\right)\right]^P
\end{align}
where $\mathrm{erfc}(\cdot)=1-\mathrm{erf}(\cdot)$ is the {complementary} error function. 

Finally, the expected value of the minimum of these $P$ random variables can be found by calculating the mean of the PDF associated with this CDF. Specifically, the expected value is derived as:
\begin{align}
\mathbb{E}(\text{min}(E_1, E_2,\ldots,E_{P})) &= \int_{-\infty}^{+\infty}
{x\, \cfrac{dF_P(x)}{dx}}\,\mathrm{d}x=\cfrac{2^{0.5-P}P}{\sqrt{\pi}\sigma}\int_{-\infty}^{+\infty}
{x\, \exp{\left[-\left(x-\mu\right)^2/2\sigma^2\right]\left[\mathrm{erfc}\left(\cfrac{x-\mu}{\sqrt{2}\sigma}\right)\right]^{P-1}}\,\mathrm{d}x} \nonumber\\
\label{eq:exact_evt_mean}
\end{align}

While Eq.~(\ref{eq:exact_evt_mean}) formally expresses the expected value of the minimum energy, it lacks a simple closed-form solution for $P>5$ due to the complexity of integrating terms involving the complementary error function. However, we can obtain an excellent approximation by focusing on the median instead of the mean, leveraging the fact that for symmetric distributions like the Gaussian, the mean and median are close in value.

So, we proceed to solve for the median $x_p$ such that $F_P(x_p)=0.5$:
\begin{align}
F_P\left(x_p\right) = 0.5 = 1-2^{-P}\left[\mathrm{erfc}\left(\cfrac{x_p-\mu}{\sqrt{2}\sigma}\right)\right]^P
\end{align}
which leads to the following solution:
\begin{align}
x_p 
= \mu 
+ \sqrt{2}\,\sigma\,\mathrm{erfc}^{-1}\bigl(2^{\,\tfrac{P-1}{P}}\bigr).
\label{eq:EVTmedianFinal}
\end{align}

which is an exact expression that finds the median. However, this expression also does not reveal the relationship between $x_p$ and $P$ explicitly.  In order to get a more revealing expression, we find an asymptotic expansion of $x_p$ which leads to:
\begin{align}
x_p \approx \mu - \sigma \sqrt{\ln{\cfrac{2}{\pi(\ln{4})^2}}+\cfrac{\ln{2}}{P}+2\ln{P}-\ln{\left[\ln{\cfrac{2}{\pi(\ln{4})^2}}+\cfrac{\ln{2}}{P}+2\ln{P}\right]}}
\end{align}
when $P\to\infty$, 
\begin{align}
x_p \approx \mu - \sigma \sqrt{2\ln{P}}
\end{align}
which is a textbook result from EVT. Thus, as $P$ increases, the minimum energy decreases as $\sqrt{\ln{P}}$, which reflects how increasing the number of replicas enhances the system's ability to find lower-energy states. 

\begin{figure*}[!ht]
    \centering
    \vspace{0pt}
    \includegraphics[width=0.95\textwidth,keepaspectratio]{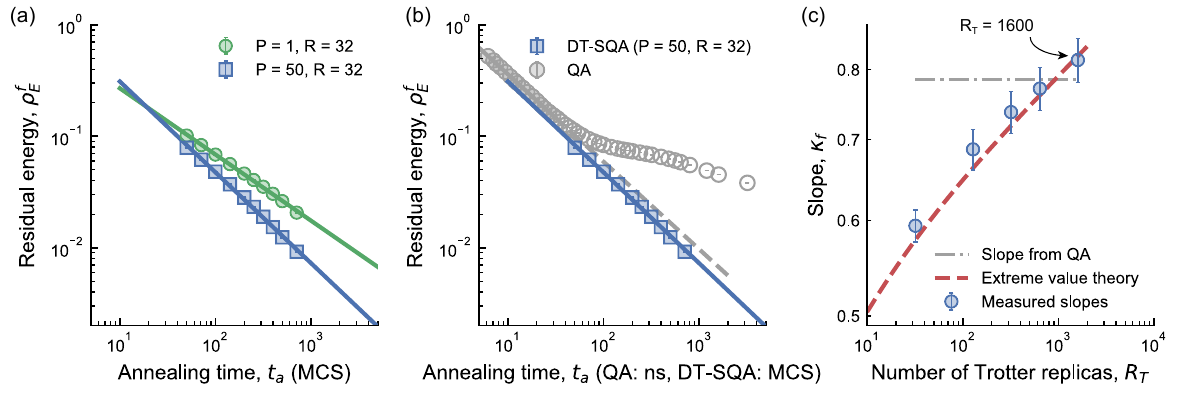}
    \caption{\footnotesize\textbf{Enhanced scaling using independent runs of Trotter replicas in DT-SQA:} (a) Residual energy ($\rho_{\mathrm{E}}^{\mathrm{f}}$) as a function of annealing time ($t_a$) is shown for $P$ independent DT-SQA simulations, each with 32 Trotter replicas. The best solution from these $P$ experiments is selected. $R_T$ is defined as the total number of Trotter replicas, and is equal to $R_T=32P$. Results are shown for $(P = 1, R=32)$ and $(P=50, R=32)$. Error bars denote the 95\% bootstrap confidence interval of the mean across spin-glass instances. (b) Comparison of DT-SQA results with $R_T=1600$ to those of the quantum annealer (QA). With  $P\approx50$, DT-SQA achieves a slope comparable to the slope reported for QA in \cite{king_coherent_2022}. (c) Slope ($\kappa_{\mathrm{f}}$) as a function of $R_T$. The red dashed line represents predictions from extreme value theory (EVT), while the gray dotted-dashed line corresponds to the slope from QA. Unlike the original DT-SQA algorithm, where all Trotter replicas are interconnected, the EVT-based approach achieves better scaling with fewer Trotter replicas. Error bars denote 95\% confidence interval of fitting. The total number of replicas $R_T$ should be compared with the total nuber of replicas $R$ in Fig.~2 in main text. Fig.~2 uses only $P=1$ which implies $R_T = R$, for simplicity we did not use $R_T$ there.} 
    \label{fig:approach2_results}
\end{figure*}

The functional form of the final residual energy, $\rho_\mathrm{E}^\mathrm{f}$ can be obtained from the Supplementary Eq.~(\ref{eq:EVTmedianFinal}) and the definition of $\rho_\mathrm{E}^\mathrm{f}$ from the main text:
\begin{align}
\rho_\mathrm{E}^\mathrm{f}(t) 
= a(t) 
+ \sqrt{2}\,b(t)\,\mathrm{erfc}^{-1}\bigl(2^{\,\frac{P-1}{P}}\bigr).
\label{eq:supp_EVT_rho}
\end{align}

where $a(t)$ and $b(t)$ are two time-dependent parameters which are related to the average mean and average standard deviation of distributions of residual energies from individual runs at a fixed MCS.  To validate our theory, we conduct $P$ independent DT-SQA simulations, each with 32 Trotter replicas, after which we select the best solution across all $P$ runs. The results of this experiment are shown in Fig.~\ref{fig:approach2_results}(a-c). We observe that with approximately $P\approx50$ repetitions, the slope achieved matches the slope reported in \cite{king_coherent_2022}.

For each MCS value, the mean ($\mu$) and standard deviation ($\sigma$) for each instance are obtained by fitting the distribution of the sampled best energies from multiple DT-SQA simulations (each simulation contains 32 Trotter replicas, and only the best energy replica is selected). The averages of these means and standard deviations are then computed across all 300 instances annealed at the same MCS. These averages are converted into residual energies using Eq.~(\ref{eq:rho_E_f}) in the main text. Residual energy predictions are derived from these mean and standard deviation values using Supplementary Eq.~(\ref{eq:supp_EVT_rho}), as shown in Fig.~\ref{fig:approach2_results}. We find excellent agreement between experimental results and predictions from conventional EVT as shown in Supplementary Fig.~\ref{fig:approach2_results}(c).

\begin{figure*}[!ht]
    \centering
    \vspace{0pt}
    \includegraphics[width=5in,keepaspectratio]{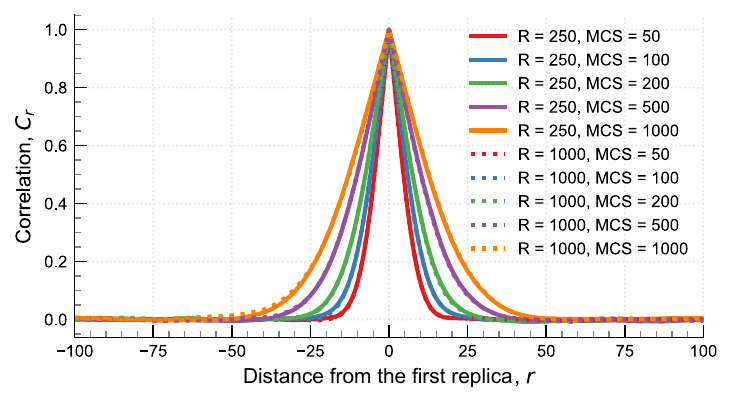}
    \caption{\footnotesize\textbf{Decay of correlation length along the replica direction:} The average correlation ($C_r = (1/n)\sum_{n}{\sigma_{i,0}\sigma_{i,r}}$) between the replicas as a function of the distance ($r$) from the first replica ($r=0$) is shown for the DT-SQA algorithm, with various total number of Trotter replicas $R$ and annealing times ($t_\mathrm{a}$, in MCS units).  We measure $C_r$ from the first replica, however, due to the periodic boundary conditions along the replica direction, $C_r$ is measured to be invariant across all replicas. The correlation length along the replica direction shows a strong dependence on the annealing times ($t_\mathrm{a}$) with longer annealing times leading to broader correlation peaks. However, the dependence on the number of Trotter replicas ($R$) is minimal, as indicated by the nearly overlapping dashed and solid lines for different $R$ values.} 
   \label{fig:corr_study}
\end{figure*}

\textbf{Modified EVT:} The extreme value theory (EVT) predictions for Fig.~\ref{fig:scaling_SQA}(c) are more involved and not exact. Two technical challenges arise: (1) all Trotter replicas are interconnected, so the block size is not known a priori, unlike the conventional EVT approach and (2) the block size may depend on the MCS values through $\Gamma_x$ and $J_{\perp}$. The latter stems from the fact that the correlation length along the replica direction varies with MCS, as shown in Fig.~\ref{fig:corr_study}. Despite these challenges, we apply EVT for an approximate understanding. We formulate a self-consistent approach as follows: we start by guessing a block size and partitioning the total number of Trotter replicas accordingly. For each partition, the best energy among the Trotter replicas within that partition is determined. As before, the mean and standard deviation of these best energy values are computed for each instance at a given MCS and averaged over all 300 instances and multiple runs per instance. Using these average mean and standard deviation values, we predict the residual energy with Supplementary Eq.~(\ref{eq:supp_EVT_rho}). This predicted residual energy is then compared with the actual residual energy computed using Eq.~(\ref{eq:rho_E_f}) in the main text. The procedure is repeated until a block size is found where the prediction and actual residual energies closely match, as shown in Supplementary Fig.~\ref{fig:blkSize}. For low MCS values, the block sizes determined from this approach approximately align with the correlation lengths obtained independently from simulations (see Supplementary Fig.~\ref{fig:corr_study}). This alignment provides a justification for our modified EVT approach. At higher MCS values, however, the distributions deviate from a Gaussian shape, becoming increasingly skewed to the left. Consequently, deviations from Supplementary Eq.~(\ref{eq:supp_EVT_rho}) become more pronounced.

\begin{figure}[!ht]
    \centering
    \vspace{0pt}
    \includegraphics[width=4.0in,keepaspectratio]{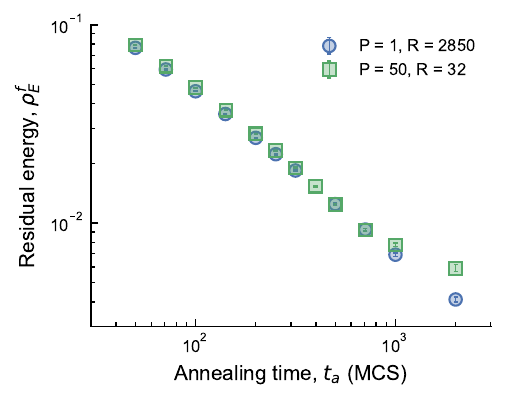}
    \caption{\footnotesize\textbf{Comparison of two DT-SQA approaches used in this work:} For logical instances of size $15\times15\times12$, EVT based approach of DT-SQA ($R 
 = 32$, $P = 50$) is compared against the conventional DT-SQA approach with $R = 2850$, $P = 1$. Both approaches share the same initial slope, which exceeds the quantum annealer's slope. However, the EVT-based approach reaches a flat plateau at a relatively higher residual energy emphasizing that the two approaches are not identical (see text).}
    \label{fig:EVT_orig_comp}
\end{figure}

\textbf{Comparison between conventional and modified EVT:} Despite the seeming similarities, the independent (conventional) and interconnected (modified) EVT approaches are not equivalent. Supplementary Fig.~\ref{fig:EVT_orig_comp} shows that the conventional approach deviates from the power-law behavior, encountering an early flat plateau around 1000 MCS. This behavior aligns well with the correlation trends shown in Supplementary Fig.~\ref{fig:corr_study}, where at around 1000 MCS the replica-to-replica correlation length exceeds the system size ($R=32$) for the conventional EVT. 

\begin{figure*}[!ht]
    \centering
    \vspace{0pt}
    \begin{minipage}{0.48\textwidth}
    \includegraphics[width=3.5in,keepaspectratio]{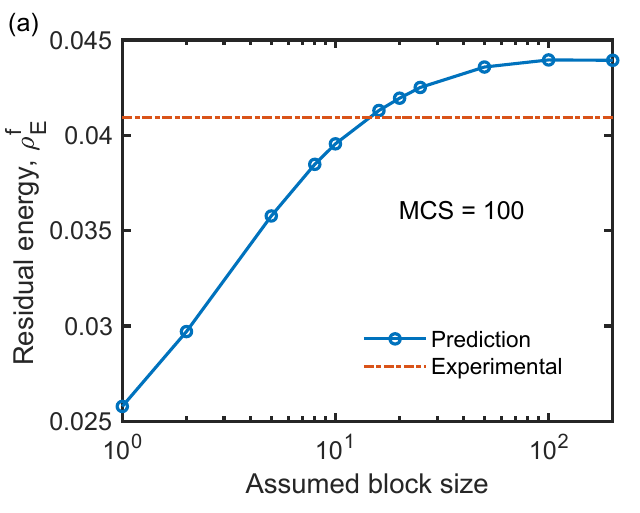}
    \end{minipage}
    \hfill
    \begin{minipage}{0.48\textwidth}
    \includegraphics[width=3.5in,keepaspectratio]{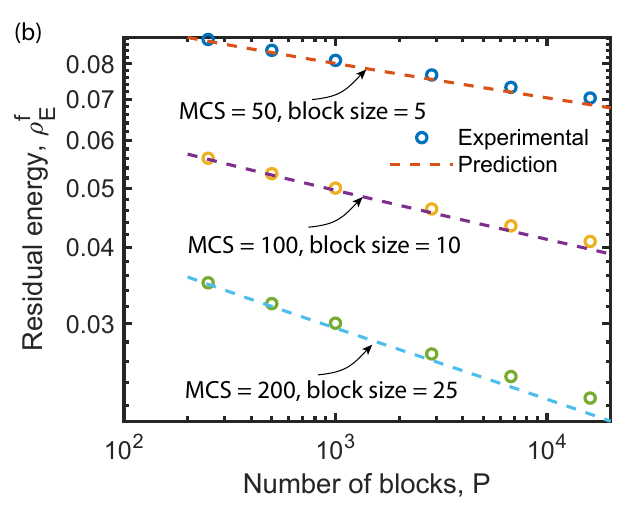}
    \end{minipage}
    \caption{\footnotesize\textbf{ Comparison of predictions from the modified EVT with experimentally obtained residual energies:} (a) The block size extraction procedure based on extreme value theory (EVT) is illustrated for $t_\mathrm{a}$ = 100 MCS. Various block sizes are assumed, and the corresponding residual energies are predicted. The block size that yields a prediction closest to the experimental residual energy is selected. (b) The procedure in (a) is repeated for different MCS values (such as $t_\mathrm{a}$ = 50, 100 and 200 MCS). The chosen block sizes for each MCS value are shown alongside the number of blocks ($P$). The predictions from the modified EVT align closely with the experimental residual energies. These results are in approximate agreement with the correlation lengths shown in Supplementary Fig.~\ref{fig:corr_study}.} 
    \label{fig:blkSize}
\end{figure*}

\section{Feasibility analysis of DT-SQA}
\label{sec:supp_chip}

\begin{figure*}[!ht]
    \centering
    \vspace{0pt}
    \includegraphics[width=0.95\textwidth,keepaspectratio]{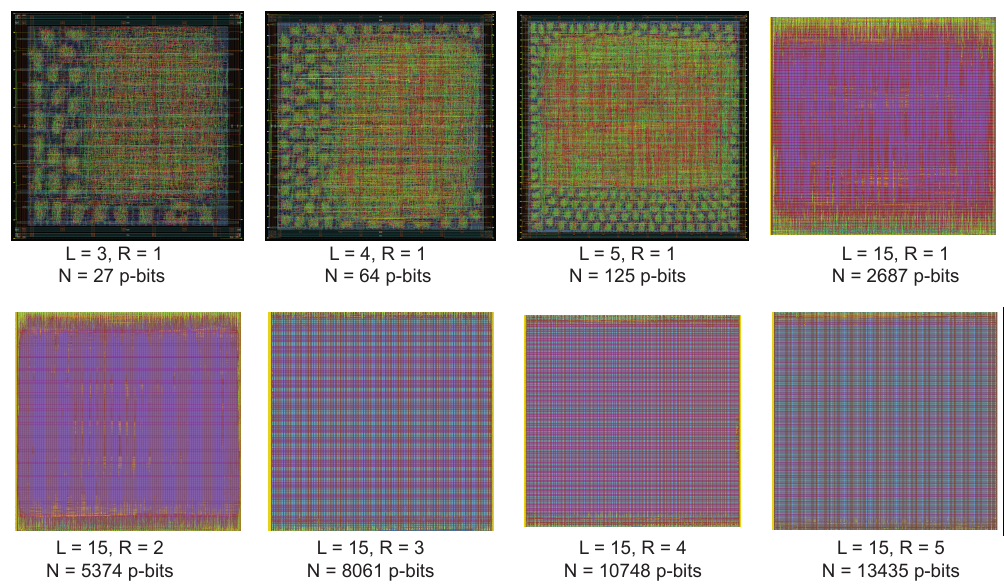}
    \caption{\footnotesize\textbf{Physical design flow results for the DT-SQA algorithm at various scales:} The feasibility of ASIC implementation for the DT-SQA algorithm is evaluated by running the mflowgen physical design flow with the ASAP7 7 nm PDK on custom RTL designs. Results are shown for different combinations of cube size ($L$) and Trotter replicas $(R)$.}
    \label{fig:GDSall}
\end{figure*}

As shown in Fig.~\ref{fig:scaling_SQA} of the main text, a large number of replicas ($R = 2850$) is required for the DT-SQA algorithm to match the scaling exponent of the quantum annealer for 3D spin glass problems. We now evaluate the feasibility of having 2850 replicas on a single chip. This way, a hardware implementation of DT-SQA would physically house and update all replicas in parallel, unlike software-based simulations where replicas are updated sequentially on a CPU.  The feasibility analysis we present here is equally applicable to the APT algorithm, which achieves performance on par with or exceeding DT-SQA and the quantum annealer while requiring fewer replicas (132 replicas as discussed in the main text) and further improvement in performance is expected with 2850 replicas. To assess the feasibility, we conduct a detailed physical design analysis. 
\begin{figure*}[!b]
    \centering
    \vspace{0pt}
    \includegraphics[width=0.95\textwidth,keepaspectratio]{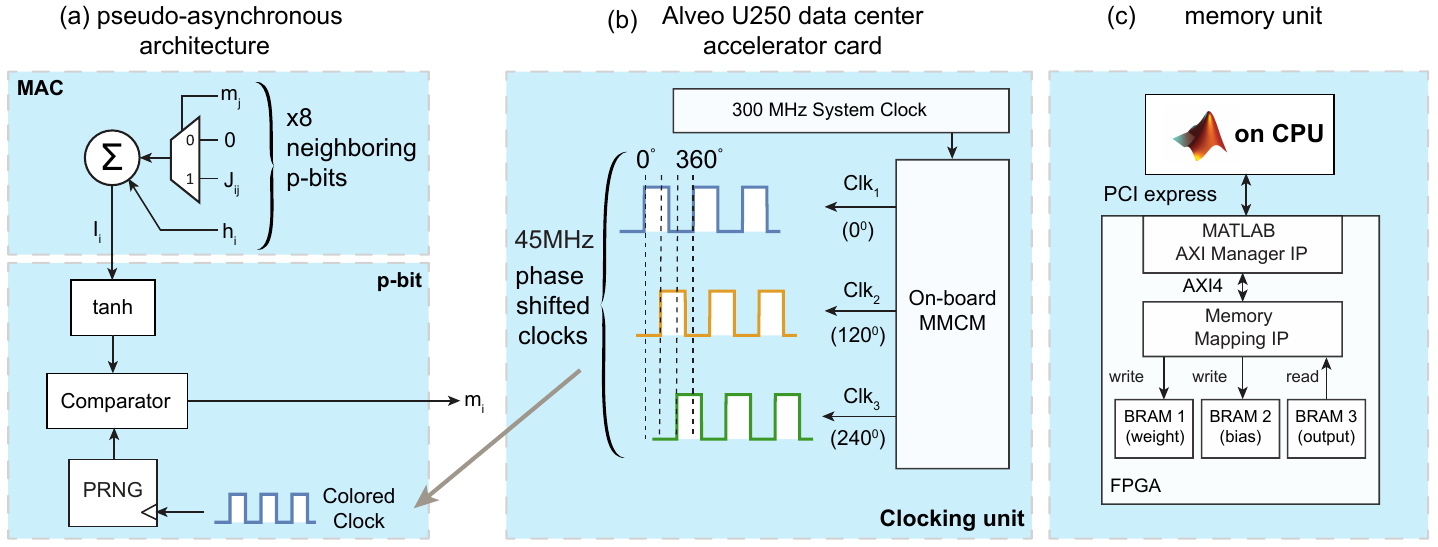}
    \caption{\footnotesize\textbf{p-computing architecture used in the feasibility analysis of the DT-SQA algorithm:}  (a) The pseudo-asynchronous architecture features a multiplier-accumulator (MAC) unit that implements Eq.~(\ref{eq:synapse}). Each p-bit unit consists of a linear feedback shift register (LFSR)-based pseudorandom number generator (PRNG), a lookup table for the activation function (tanh), and a comparator to generate a binary output. (b) The clocking unit on the Alveo U250 data center accelerator card generates three phase-shifted clocks (0°, 120°, and 240°) at 45 MHz from a 300 MHz system clock using an on-board mixed-mode clock manager (MMCM). These are used to trigger the PRNGs inside the colored p-bit blocks. (c) The memory unit interfaces with the CPU via Peripheral Component Interconnect (PCI) Express. Data transfer between MATLAB and the FPGA is managed through Advanced eXtensible Interface (AXI) interfaces, with BRAMs (block RAMs) allocated for weights, biases, and binary p-bit outputs.  Weights and biases have  fixed point  s\{6\}\{3\} precision  where `s' denotes the sign bit and the first and second curly braces represent the integer and fractional parts.
 }
    \label{fig:FPGA_arch}
\end{figure*}
For our analysis, we use the open-source mflowgen physical design flow \cite{mflowgen} with the open-source ASAP7 7 nm process design kit (PDK) \cite{asap7}. Custom RTLs are developed to implement the algorithm based on the p-computing architecture shown in Supplementary Fig.~\ref{fig:FPGA_arch}, incorporating multi-phase clocking to manage timing across the design. Synthesis is carried out using Cadence Genus, followed by floorplanning, power distribution network (PDN) generation, clock tree synthesis (CTS), and final place-and-route (P\&R) using Cadence Innovus. Fig.~\ref{fig:GDSall} provides examples of designs after placement and routing, with the corresponding design metrics summarized in Table~\ref{tab:mflowgen_results}. The routed designs are verified to meet timing constraints under a 3-phase clock (to allow provision for using odd number of replicas), with each phase operating at a frequency of 100 MHz. The final designs, completed after placement and routing, demonstrate that: (1) an ASIC implementation is feasible, and
(2) the area required for this ASIC scales linearly with problem sizes. 

\begin{figure*}[!ht]
    \centering
    \vspace{0pt}
    \includegraphics[width=4.5in,keepaspectratio]{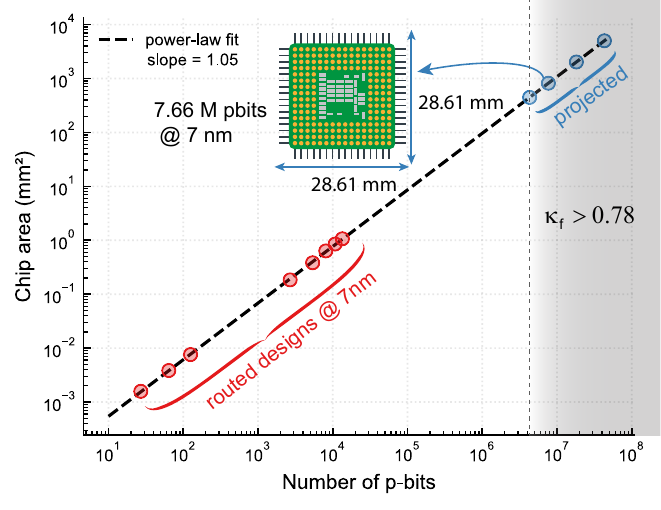}
    \caption{\footnotesize\textbf{Scaling of chip area with the number of p-bits for the DT-SQA algorithm:} The growth in chip area is studied as more p-bits are integrated into the chip, following a full place-and-route design process. The observed trend exhibits near-linear growth, with a slope of 1.05. The largest routed designs, based on the 7 nm ASAP7 PDK, are shown in red, while the projected scaling is represented in blue. Extrapolating to a modern chip size of  $28.61\times28.61$ mm\textsuperscript{2}, it is estimated that such a chip could accommodate approximately 7.66 million p-bits. This corresponds to 2850 replicas of size $15\times15\times12$, achieving scaling comparable to a quantum annealer with $\kappa_{\mathrm{f}}>0.785$.}
    \label{fig:chip_area}
\end{figure*}

Fig.~\ref{fig:chip_area} shows the area scaling of placed and routed designs for different instances of the DT-SQA algorithm. The area scaling follows a linear trend, with the largest design--featuring 13435 p-bits--occupying approximately 1 mm\textsuperscript{2}. Extrapolating this scaling to modern chip dimensions, a chip measuring 28.61 mm $\times$ 28.61 mm could accommodate approximately 7.66 million p-bits, corresponding to 2850 replicas of size $15\times15\times12$, achieving a scaling similar to that of a quantum annealer.

\begin{table}[!ht]
\vspace{0pt}
\centering
\caption{\textbf{Details of the physical flow designs:} Detailed information about the physical flow design performed in this study are listed. The design results are based on ASAP7 - a 7 nm finFET predictive process design kit (PDK) with mflowgen. $L$ represents the dimension of the cube and $R$ denotes the number of replicas used.}
\vspace{0.2cm}
\begin{tabular}{@{}LGGGGGGGGGG@{}}
\toprule
\textbf{Number of p-bits} & 27 & 64 & 125 & 2687 & 5374 & 8061 & 10748 & 13435 & $7.66 \times10^6$\\ \midrule
\textbf{Cube dimension}, $L$ & 3 & 4 & 5 & 15 &  15 & 15 & 15 & 15 & 15\\
\textbf{Number of replicas}, $R$ & 1 & 1 & 1 & 1  & 2 & 3 & 4 & 5 & 2850\\
\textbf{Flow status} & routed & routed & routed & routed & routed & routed & routed & routed& projected\\
\textbf{Number of standard cells} & 11303 & 27653 &  54685  & 1.385$\times10^6$  & 2.978$\times10^6$  & 4.886$\times10^6$  & 6.562$\times10^6$  & 8.199$\times10^6$ & $4.7\times10^9$\\
\textbf{Timing met (per phase) MHz} & 100 & 100 & 100 & 100 & 100 & 100 & 100 & 100 & 100 \\
\textbf{Area (mm\textsuperscript{2})} & 0.001581 & 0.003844 & 0.007594 & 0.184317 & 0.382561 & 0.634809 & 0.851805 & 1.065527 & 818.34\\
\bottomrule
\label{tab:mflowgen_results}
\end{tabular}
\end{table}

\section{APT Algorithm with Isoenergetic Cluster Move (ICM)}
\label{supp_sec:APT_algo}

Parallel tempering (PT) and its variants are a standard choice for solving challenging optimization problems such as the 3D spin glass. Hence, we also evaluate the performance of PT. Like DT-SQA, PT is also a replica-based algorithm. For our analysis, we employ an adaptive version of PT (APT) that includes a preprocessing step to determine the temperature schedule and optimize the number of replicas required. To further enhance the performance of APT, we incorporate isoenergetic cluster moves (ICM). A pseudocode detailing the adaptive parallel tempering algorithm, including the temperature schedule preprocessing and ICM, is presented in Algorithm~\ref{algo:APT}. Additional details regarding the exact parameters used can be found in the Methods section of the main text.

\SetKwComment{Comment}{\normalfont $\triangleright$ }{}
\SetKwInput{KwInput}{Input}                
\SetKwInput{KwOutput}{Output}              
\SetKwFor{ParFor}{for}{do in parallel}{end}
\begin{algorithm}[!ht]
    \DontPrintSemicolon
    \KwInput{Weights, biases, number of swaps,  sweeps per swap, colormap, step rate $\alpha$, initial inverse temperature $\beta_0$, energy variance tolerance, number of chains, sweeps per chain}
    
    \KwOutput{State corresponding to minimum energy, ${m}_{\mathrm{opt}}$}
    \SetKwFunction{FMain}{p-computer}
    \SetKwProg{Fn}{Function}{:}{}
    \Fn{\FMain{weights, biases, colormap, temp.}}{
        \For{each color in the colormap}{
            \For{each p-bit in the color}{
                Solve Eq.~(\ref{eq:synapse}) and Eq.~(\ref{eq:neuron}).\;
            }
        }
    }
    
    \SetKwFunction{Fcluster}{ICMop}
    \SetKwProg{Fnew}{Function}{:}{}
    \Fnew{\Fcluster{replica1, replica2}}{
        Find the overlap vector between replica 1 and replica 2.\;
        Randomly pick one cluster where overlap is $-1$.\;
        \If{size of the cluster is greater than half of the total number of spins}{
            Randomly chose one of the two replicas.\;
            Flip all the spins of the chosen replica.\;
        }
        \Else{
            Flip all the spins inside the chosen cluster of replica 1 and replica 2.\;
        }
    \KwRet replica 1, replica 2
    }

   $t\gets 0$, $\beta_t \gets \beta_0$\\
   Initialize all parallel chains to random states.\; 
  
  \While{energy variance is greater than tolerance}
    {
        \For{each chain in parallel}{
            \For{each sweep}{
            Sample p-bit states from p-computer.\;
            Compute energy of the chain.\;
            }
            Compute energy variance for the chain.\;
            Save the p-bit states.\;
            }
        Compute mean energy variance of chains, $\sigma_{\text{E}}$.\; 
        Set next step inverse temperature: $\beta_{t+1} \gets \beta_t + \displaystyle{\frac{\alpha}{\sigma_{\text{E}}}}$,\,\,$t\gets t+1$.\;
     }
    
    Initialize all replicas to random states\\
    \For{each swap attempt}{
        \If{it is an even numbered swap attempt}{
            Choose (even, odd) sequential pairs.\;
        }
         \Else{
            Choose (odd, even) sequential pairs.\;
        }
        \For{each replica in parallel do}{
            \For{each sweep}{
                 Sample p-bit states from p-computer.\;       
            }
            Compute energy of the replica.\;
            Randomly partition ICM replicas into pairs.\;
            \For{each ICM replica pair}{
                Perform ICMop on the replicas in the pair.\;
            } 
        }
        \For{each sequential pair of replicas}{
            Propose a swap.\;
            \If{accepted}{
                Swap the p-bit states of all corresponding iso-temperature replicas between two replicas.\;
            }
        }
    }
    \KwRet p-bit states for the replica with the minimum energy.\;
\caption{Adaptive parallel tempering with p-computers}
\label{algo:APT}
\end{algorithm}

\subsection{Temperature profile and swap acceptance rate in APT}
\label{supp:beta_profiles}

\begin{figure}[!ht]
    \centering
    \vspace{0pt}
    \begin{minipage}{0.48\textwidth}
    \includegraphics[width=3.5in,keepaspectratio]{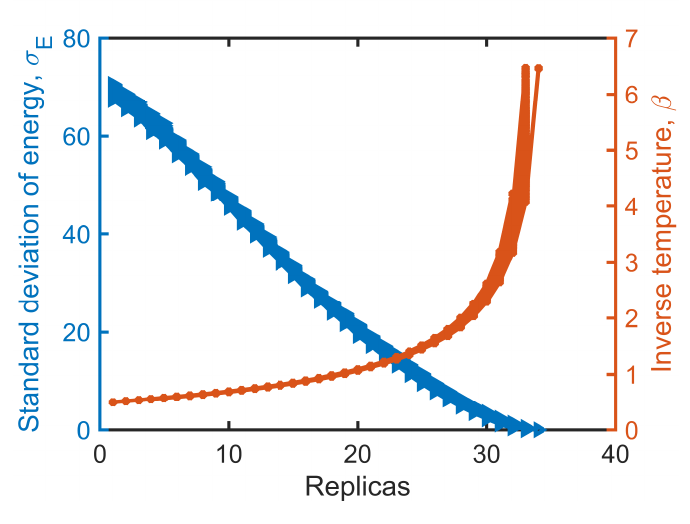}
    \caption{\footnotesize\textbf{Inverse temperature profile of the APT algorithm:} The inverse temperature ($\beta$) profiles generated by the preprocessing algorithm for each of the 300 instances are shown. At high temperatures (low $\beta$), the profiles across instances are nearly identical, with slight deviations appearing at low temperatures (high  $\beta$). The figure also shows the standard deviation of sampled energies ($\sigma_{\mathrm{E}}$) at each temperature, which decreases monotonically as $\beta$ increases. The preprocessing begins with $\beta=0.5$, below which the standard deviation saturates to a constant value. The preprocessing step terminates when the average standard deviation of energy drops below the minimum coupling value, min($J_{ij}$).}
    \label{fig:replica_profile}
    \end{minipage}
    \hfill
    \begin{minipage}{0.48\textwidth}
    \centering
    \vspace{-20pt}
    \includegraphics[width=3.5in,keepaspectratio]{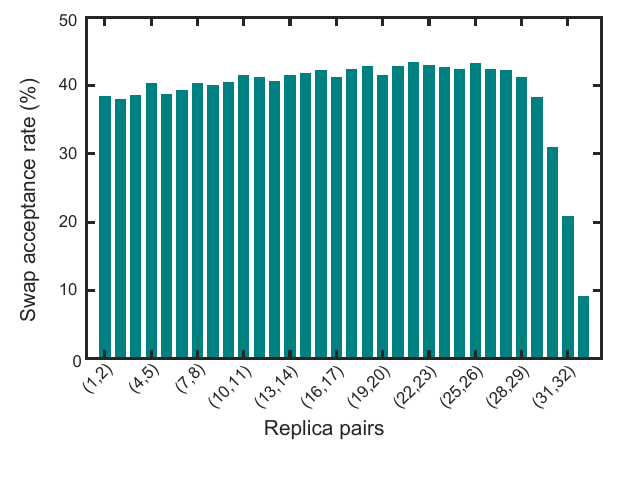}
    \vspace{-15pt}
    \caption{\footnotesize\textbf{Swap acceptance rate of the APT algorithm:} The swap acceptance rate obtained from the APT with ICM algorithm for a randomly chosen instance using the preprocessed temperature profiles is shown. With $\alpha = 1.25$, the average swap acceptance rate across all nearest-neighbor replica pairs is approximately 40\%. The results are averaged over 10000 swaps and six independent runs. 4 ICM replicas are used, but their averages are reported as a single replica.}
    \label{fig:swap_rate}
    \end{minipage}
\end{figure}

We apply the preprocessing algorithm individually to each problem instance as detailed in the Methods section. The temperature profiles for all 300 instances of size $15\times15\times12$ are shown in Fig.~\ref{fig:replica_profile}. The profiles are highly consistent across instances, with only slight deviations in the low-temperature (high $\beta$) region. In principle, a temperature profile generated for a randomly selected instance could be applied to all instances without significantly impacting performance. However, in this work, we optimize the temperature profiles for each instance.

Additionally, Fig.~\ref{fig:replica_profile} includes the standard deviation of the sampled Monte Carlo energies at each iteration (corresponding to each $\beta$). The standard deviation decreases monotonically as $\beta$ increases, confirming that the replica temperatures are chosen appropriately.

Fig.~\ref{fig:swap_rate} shows the typical swap acceptance rate in the APT algorithm for logical instances of size 
$15\times15\times12$. For optimal performance, it is generally recommended that the acceptance rate remains approximately constant. As shown, the acceptance rate stays roughly constant at around 40\%, except for the last few replicas. However, since the algorithm selects the replica with the minimum energy at each swap, the lower acceptance rate of the last few replicas does not impact the overall performance of the algorithm.

\vspace{-5pt}
\subsection{Improvement in the performance of APT algorithm with ICM}
\begin{figure*}[!ht]
    \centering
    \vspace{0pt}
    \includegraphics[width=7.0in,keepaspectratio]{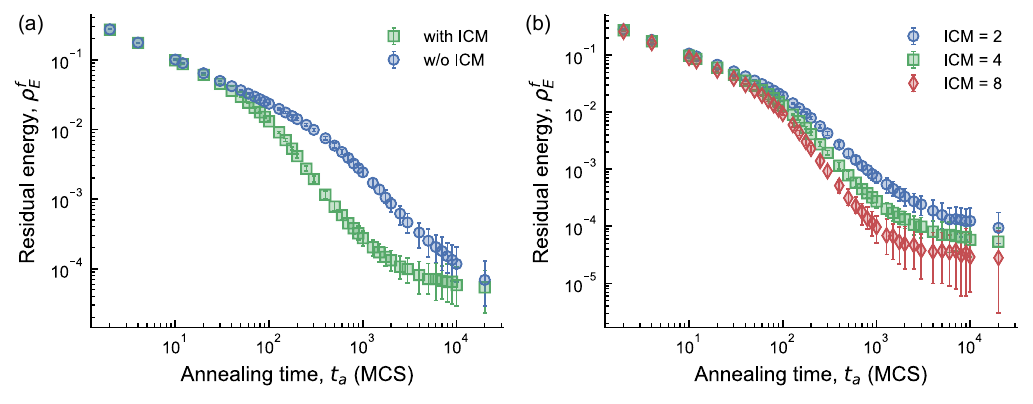}
    \caption{\footnotesize\textbf{Residual energy ($\rho_\mathrm{E}^\mathrm{f}$) from APT with and without ICM:} (a) The residual energy as a function of annealing time ($t_\mathrm{a}$) is shown for the APT algorithm, both with and without isoenergetic cluster moves (ICM) and for cube size $L = 8$ ($512$ spins). The results demonstrate that incorporating ICM improves performance, yielding lower residual energy compared to APT without ICM, particularly at longer annealing times. In this analysis, one Monte Carlo sweep is performed for each replica before a swap is attempted and 4 ICM replicas are used. (b) The impact of varying the number of ICM replicas is illustrated, with performance improving as the number of ICM replicas increases.}
    \label{fig:APTvsICM}
\end{figure*}

Here, we justify the inclusion of ICM and compare the performance of APT with and without ICM, as shown in Fig.~\ref{fig:APTvsICM}(a). Despite the additional computational overhead introduced by ICM, it offers several advantages: (1) It achieves lower residual energy compared to APT without ICM for a fixed MCS budget, with the difference becoming more pronounced at larger $t_\mathrm{a}$ values. (2) It delivers an improved slope compared to APT without ICM. Adding more ICM replicas improves the performance of the algorithm as shown in Fig.~\ref{fig:APTvsICM}(b).

\begin{figure}[!ht]
    \centering
    \vspace{0pt}
    \includegraphics[width=4.0in,keepaspectratio]{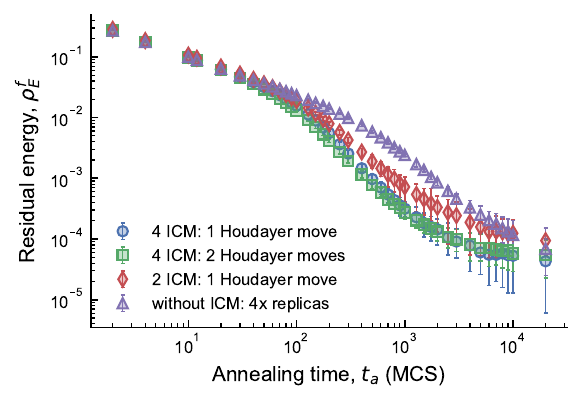}
    \caption{\footnotesize\textbf{Effect of ICM on residual energy ($\rho_\mathrm{E}^\mathrm{f}$) in APT with ICM algorithm:}  The performance of APT with ICM is compared with varying ICM per MCS to investigate the role of ICM. A Houdayer move \cite{Houdayer2001} is an isoenergetic cluster update used in spin-glass Monte Carlo simulations. It operates by identifying a cluster of antiparallel spins between two replicas and swapping them to enhance mixing and improve sampling efficiency. With 4 replicas for ICM, two replica pairs are available for ICM--one Houdayer move on each pair. The purple triangles represent APT without any ICM. The blue circles use one ICM on a randomly chosen pair, and the green squares use two ICMs. These three plots use the same number of replicas for a fair comparison. For comparison, we also show APT with 2 ICM replicas (red diamonds) which has only one replica pair for ICM.}
    \label{fig:Houdayer_effect}
\end{figure}

A logical question that arises is whether the bending observed in APT with ICM is primarily due to the increased number of replicas or the inclusion of ICM. Supplementary Fig.~\ref{fig:Houdayer_effect} addresses this by comparing APT with and without ICM. The impact of ICM is evident from the clear separation between the cases: APT using ICM (blue circles and green squares) show lower residual energy compared to those without ICM (purple triangles), even for the same number of replicas. Additionally, performance improvements are observed as the number of replicas increases, further contributing to the bending. This emphasizes the critical role of non-local moves in classical/probabilistic algorithms, which can significantly enhance performance. A natural next step could be to use non-equilibrium Monte Carlo algorithms aimed to improve the APT algorithm \cite{Mohseni2021}, but we do not attempt this here. 

\vspace{-5pt}
\subsection{APT with ICM as a function of sweep to swap ratio}
Next, we evaluate the performance of APT with ICM as a function of the sweep-to-swap ratio, which defines the number of sweeps performed before each swap attempt. Our findings indicate that the sweep-to-swap ratio significantly impacts the algorithm performance. While a lower sweep-to-swap ratio increases the number of swaps, it consistently results in better residual energy for a fixed MCS budget, as shown in Fig.~\ref{fig:ICM_sweep_swap} for three different sweep-to-swap ratios. Replica energies are calculated at each swap attempt and not saved for the entire annealing time. The sweep-to-swap ratio of 1 gives the best residual energy for an MCS budget even though a more typical sweep-to-swap ratio of 10 achieves close performance with a similar bending behavior, therefore this sweep-to-swap choice does not critically change our results. We carefully verified that computing replica energy at each sweep does not affect the conclusion, confirming that the superior performance at a sweep-to-swap ratio of 1 is not simply a result of increased computational effort.  The additional computational cost from more frequent swaps can be mitigated using dedicated hardware, as described in the Methods section of the main text.

\subsection{Performance of APT with ICM on the embedded instances}
\begin{figure*}[!ht]
    \centering
    \vspace{0pt}
    \begin{minipage}{0.45\textwidth}
    \includegraphics[width=3.5in,keepaspectratio]{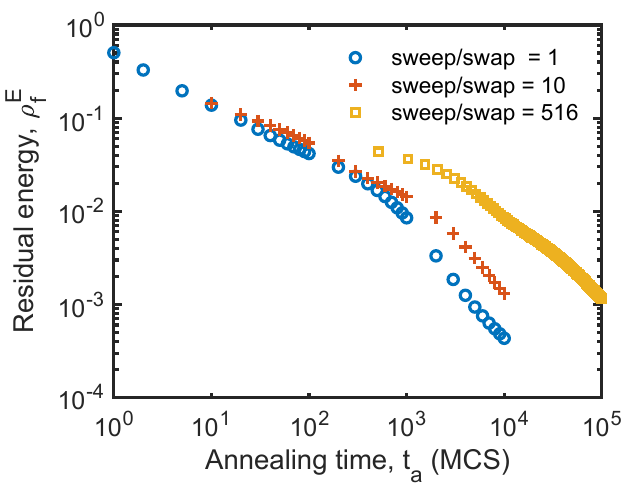}
    \caption{\footnotesize\textbf{Residual energy ($\rho_\mathrm{E}^\mathrm{f}$) with APT + ICM as a function of sweep-to-swap ratio:}  The performance of the APT with ICM algorithm is evaluated for various sweep-to-swap ratios (defined as the number of Monte carlo sweeps performed for each replica before a swap is attempted). The residual energy is plotted as a function of annealing time ( $t_\mathrm{a}$) for sweep-to-swap ratios of 1, 10, and 516. The results show that a lower sweep-to-swap ratio (sweep/swap = 1) yields the best performance, achieving lower residual energy and a steeper slope, indicating faster convergence towards solutions. In this analysis, 4 ICM replicas are used for ICM.}
    \label{fig:ICM_sweep_swap}
    \end{minipage}
    \hfill
    \begin{minipage}{0.44\textwidth}
    \includegraphics[width=3.5in,keepaspectratio]{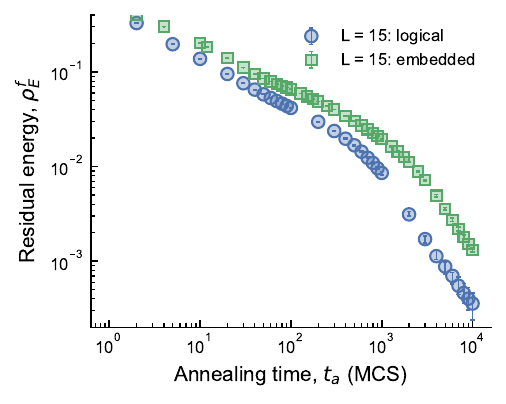}
    \caption{\footnotesize\textbf{Residual energy ($\rho_\mathrm{E}^\mathrm{f}$) with APT + ICM for embedded instances:} The performance of the APT with ICM algorithm is shown for embedded instances. The residual energy is plotted as a function of annealing time ( $t_\mathrm{a}$) for sweep-to-swap ratios of 1 and 4 ICM replicas compared with logical instances of the same size ($L = 15$). The embedded instances show similar characteristics to those of logical instances.}
    \label{fig:APT_embedded}
    \end{minipage}
\end{figure*}

Fig.~\ref{fig:APT_embedded} compares the performance of APT with ICM on the embedded instances. The performance is very similar to the logical instances showing the transition from a gentler to steeper slope.

\begin{figure}[!ht]
    \centering
    \vspace{0pt}
    \includegraphics[width=0.925\textwidth,keepaspectratio]{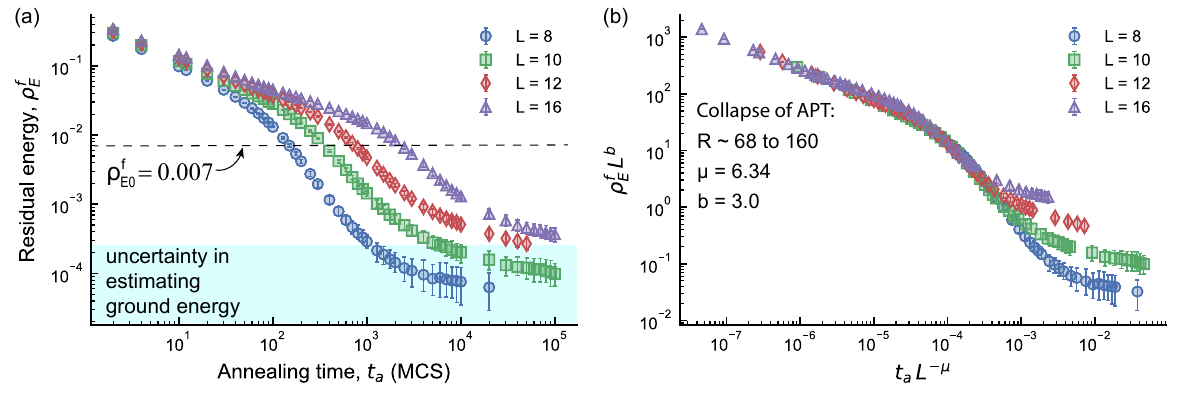}
    \vspace{0pt}\caption{\footnotesize\textbf{Residual energy vs. annealing time $t_a$ plots as a function of system size $L$ of the logical instances for APT with ICM:} (a) Residual energies plotted against the annealing time for a few system sizes $L$.  All data points are averaged over 100 initial conditions per each of the 300 instances provided/generated. (b) The collapse of the residual energy versus annealing time is shown, using the finite size scaling method. Setting $\mu = 6.34$, and  $b=3$ provides a gentle collapse of the data onto a single universal curve although the collapse breaks down at very low residual energies near the ground states of the instances, probably due to the uncertainty in the determination of the ground energy.  The $L = 15$ data is excluded from this collapse because it does not correspond to an exact cube.}
    \label{fig:rho_ta_L}
\end{figure}

\begin{figure}[!ht]
    \centering
    \vspace{0pt}
    \includegraphics[width=0.95\textwidth,keepaspectratio]{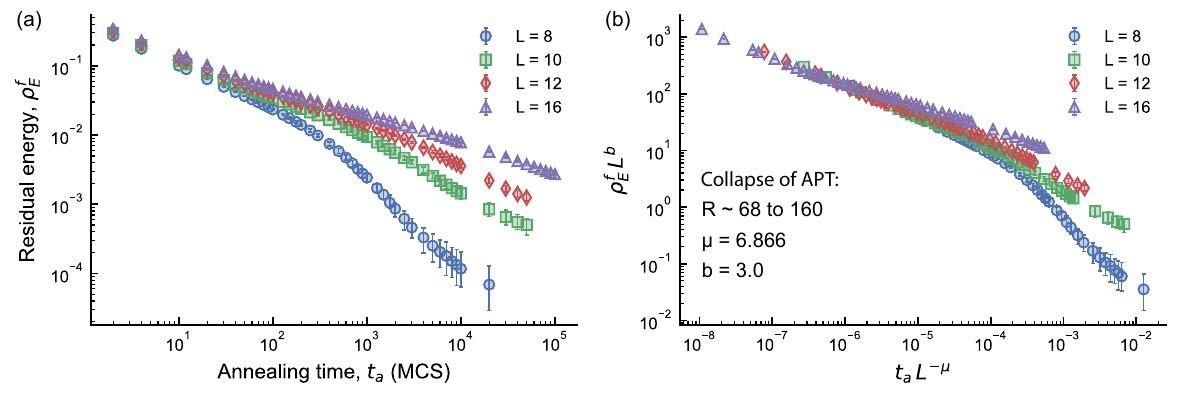}
    \vspace{0pt}\caption{\footnotesize\textbf{Residual energy vs. annealing time $t_a$ plots as a function of system size $L$ of the logical instances for APT without ICM:} (a) Residual energies plotted against the annealing time for a few system sizes $L$.  All data points are averaged over 100 initial conditions per each of the 150 instances provided/generated. (b) The collapse of the residual energy versus annealing time is shown, using the finite size scaling method. Setting $\mu = 6.866$, and  $b=3$ provides a gentle collapse of the early part of the data onto a single universal curve. The later part of the data does not show a second bending like the universal curve for the APT with ICM algorithm.}
    \label{fig:rho_ta_L_woICM}
\end{figure}

\vspace{-5pt}
\subsection{Slope of the residual energy as a function of system size of the logical instances}

In Supplementary Fig.~\ref{fig:rho_ta_L}, we show the final residual energy as a function of annealing time for four different system sizes, for the APT with ICM algorithm. We perform a scaling analysis: the annealing time is rescaled as 
$t_\mathrm{a} L^{-\mu}$, and the residual energy is rescaled as $L^{b}\rho_{\mathrm{E}}^{\mathrm{f}}$. This rescaling collapses the data onto a single universal curve, indicating that the system behavior follows a universal finite-size scaling law. The $L=15$ data is excluded from the collapse analysis because these instances do not form a perfect cube ($15\times 15\times 12$).  At very low residual energies we also observe another transition to a gentler slope, probably due to the uncertainty in the determination of the ground energy (see Methods section of the main text). We define an arbitrarily chosen target residual energy, $\rho_{\mathrm{E}0}^f$, to approximate the optimization performance. For larger system sizes, achieving a given residual energy target requires progressively longer annealing times. The observed universal collapse confirms that the annealing time needed to reach any target residual energy can be predicted approximately for any system size. Supplementary Fig.~\ref{fig:rho_ta_L_woICM} shows similar plots to those in Supplementary Fig.~\ref{fig:rho_ta_L}, but for the APT algorithm without ICM. To ensure a fair comparison, the number of replicas is kept consistent with that used in the APT with ICM algorithm. Unlike the case with ICM, the collapse here does not exhibit a second bending.

\vspace{-5pt}

\section{FPGA implementation of Adaptive Parallel Tempering}
\label{supp_sec:FPGA_verif}

\begin{figure}[!ht]
    \centering
    \vspace{0pt}
    \includegraphics[width=0.48\linewidth,keepaspectratio]{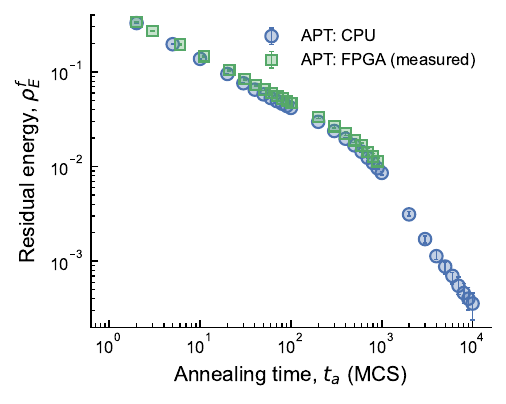}
    \vspace{-10pt}\caption{\footnotesize\textbf{Verification of APT with ICM algorithm implemented on FPGA:}  Adaptive parallel tempering (APT) with isoenergetic cluster moves (ICM) on CPUs for the problem size $15\times15\times12$, averaging over 300 instances and using 50 initial conditions per instance. A sweep-to-swap ratio of 1 minimizes residual energy for APT. Also shown FPGA implementation of APT with ICM, running 10 instances with 10 initial conditions each, closely matching the CPU result. Deviations between the FPGA and CPU may be due to the fixed point weight precision used in the FPGA (s\{6\}\{6\}, where `s' denotes the sign bit and the first and second curly braces represent the integer and fractional part of the weights, respectively) compared to float64 in CPU and possibly due to the pseudorandom number generator differences (LFSR in the FPGA and Mersenne Twister in the CPU), though this difference typically does not play a significant role. Note that this precision is higher than what we used for DT-SQA in our feasibility analysis, due to the sensitivity of the APT algorithm to the weight precision.  Error bars denote 95\% bootstrapped confidence interval of mean over instances.}
    \label{fig:APT_FPGA_data}
\end{figure}

To evaluate hardware acceleration, we implemented the APT with ICM algorithm on a moderately sized FPGA capable of supporting approximately 5000 p-bits. For system size $L=15$, the FPGA results (Supplementary Fig.~\ref{fig:APT_FPGA_data}) closely match CPU simulations, verifying the correctness of the implementation. The details of the implementation are provided in the Methods section. At this scale, implementation of APT without ICM requires 32 to 34 replicas, while APT with ICM requires 128 to 136 replicas. To overcome FPGA resource constraints, we employ time-division multiplexing (TDM), allowing the same hardware to be reused for multiple replicas. However, this introduces communication overhead, primarily from saving and reloading p-bit states and performing off-chip energy calculations. As a result, FPGA experiments are limited to 10 instances, 10 runs, and a maximum of 1000 MCS. We emphasize that this is not a fundamental limitation. As our hardware feasibility analysis with custom integrated circuits shows, larger FPGAs or ASICs could eliminate the need for TDM. In addition, energy calculations required for the PT swaps can be performed on chip,  reducing overhead and improving scalability. Currently, our FPGA setup can accommodate only one replica of size $15\times15\times12$ (2687 p-bits) or $16\times16\times16$ (4096 p-bits). The architecture utilizes graph coloring to maximize parallelism, achieving one sweep per replica in 22.22 ns (45 MHz clock; see Methods Section), corresponding to 120 and 185 flips per ns for the respective problem sizes. Performance can be further improved by increasing the number of on-chip p-bits or interconnecting multiple chips.  Our implementation achieves a flips per ns metric 50 to 75 times higher than the 2.5 flips per ns reported for optimized simulated annealing on CPUs \cite{King2023quantum}.

\end{document}